\documentclass[sigconf]{acmart}

\usepackage{graphicx}
\usepackage{multirow}
\usepackage{makecell}

\usepackage{tikz}
\usepackage{amsmath}
\usepackage{amssymb}

\newcommand*\circled[1]{\tikz[baseline=(char.base)]{
		\node[shape=circle,draw,inner sep=0.8pt] (char) {#1};}}

\AtBeginDocument{%
  \providecommand\BibTeX{{%
    \normalfont B\kern-0.5em{\scshape i\kern-0.1em b}\kern-0.1em\TeX}}}

\copyrightyear{2020}
\acmYear{2020}
\setcopyright{acmcopyright}
\acmConference[MSR '20]{17th International Conference on Mining Software Repositories}{October 5--6, 2020}{Seoul, Republic of Korea}
\acmBooktitle{17th International Conference on Mining Software Repositories (MSR '20), October 5--6, 2020, Seoul, Republic of Korea}
\acmPrice{15.00}
\acmDOI{10.1145/3379597.3387449}
\acmISBN{978-1-4503-7517-7/20/05}

\begin{document}

\title{Improved Automatic Summarization of Subroutines via Attention to File Context}

\author{Sakib Haque}
\email{shaque@nd.edu}
\affiliation{%
  \institution{Dept. of Computer Science\\University of Notre Dame}
  \city{Notre Dame}
  \state{IN}
  \country{USA}
  \postcode{46656}
}

\author{Alexander LeClair}
\email{aleclair@nd.edu}
\affiliation{%
	\institution{Dept. of Computer Science\\University of Notre Dame}
	\city{Notre Dame}
	\state{IN}
	\country{USA}
	\postcode{46656}
}

\author{Lingfei Wu}
\email{wuli@us.ibm.com}
\affiliation{%
	\institution{IBM Research}
	\city{Yorktown Heights}
	\state{NY}
	\country{USA}
	\postcode{10598}
}

\author{Collin McMillan}
\email{cmc@nd.edu}
\affiliation{%
	\institution{Dept. of Computer Science\\University of Notre Dame}
	\city{Notre Dame}
	\state{IN}
	\country{USA}
	\postcode{46656}
}

\renewcommand{\shortauthors}{Haque, et al.}

\begin{abstract}
Software documentation largely consists of short, natural language summaries of the subroutines in the software.  These summaries help programmers quickly understand what a subroutine does without having to read the source code him or herself.  The task of writing these descriptions is called ``source code summarization'' and has been a target of research for several years.  Recently, AI-based approaches have superseded older, heuristic-based approaches.  Yet, to date these AI-based approaches assume that all the content needed to predict summaries is inside subroutine itself.  This assumption limits performance because many subroutines cannot be understood without surrounding context.  In this paper, we present an approach that models the file context of subroutines (i.e. other subroutines in the same file) and uses an attention mechanism to find words and concepts to use in summaries.  We show in an experiment that our approach extends and improves several recent baselines.
\end{abstract}



\maketitle

\section{Introduction}

One of the most important aspects of software documentation is the generation of short, usually one-sentence, descriptions of the subroutines in the software source code.  These descriptions are often the backbone of documentation tools such as JavaDoc, Doxygen, or Swagger~\cite{kramer1999api}.  They are important because they help programmers navigate source code to understand what subroutines do, the role they play in a program, and how to use them~\cite{forward2002relevance}.  Even a short description such as ``initializes microphone for web conference'' says a lot to a programmer about what a subroutine does.  The task of generating these descriptions has become known as ``summarization'' of subroutines.  The problem definition is quite simple: given the source code for a subroutine, generate a one-sentence description of that subroutine.  Yet while currently a vast majority of summarization is handled manually by a programmer, automatic summarization has a long history of scientific interest and been described as a ``holy grail''~\cite{leclair2019neural} of SE research.

A flurry of recent research has started to make automatic summarization a reality.  Following the pattern in many research areas, early efforts were based on manual encoding of human knowledge such as sentence templates~\cite{sridhara2010towards, sridhara2011automatically, mcburney2016automatic}, until around 2016-2018 when AI-based, data-driven approaches superseded manual approaches.  Nearly all of the literature on these AI-based approaches to subroutine summarization is inspired by Neural Machine Translation (NMT) from the Natural Language Processing research area.  In the typical NMT problem, a neural model is trained using pairs of sentences in one language and their translation in another language.  A stereotyped application to code summarization is that pairs of code and description are used to train a model instead -- code serves as one ``language'' and descriptions as the other~\cite{iyer2016summarizing}.

These existing approaches have shown promising, but not yet excellent, performance.  The first techniques focused on an off-the-shelf application of an encoder-decoder neural architecture such as by Iyer~\emph{et al.}~\cite{iyer2016summarizing}, with advancements looking to squeeze more information from the code such as by Hu~\emph{et al.}~\cite{hu2018deep} and LeClair~\emph{et al.}~\cite{leclair2019neural} using a flattened abstract syntax tree, and very recently Alon~\emph{et al.}~\cite{alon2018code2seq} and Allamanis~\emph{et al.}~\cite{allamanis2018learning} using graph neural nets and execution paths.  Yet, all of these approaches are based on an assumption that a subroutine can be summarized using only the code inside that subroutine: the only input is the code of the subroutine itself, and the model is expected to output a summary based solely on that input.  But this assumption has led to controversy for neural-based solutions~\cite{hellendoorn2017deep} since program behavior is determined by the interactions of different subroutines, and the information needed to understand a subroutine is very often encoded in the context around a subroutine instead of inside it~\cite{hill2009automatically, mcburney2016automatic, holmes2005using, biggerstaff1993concept}.

In this paper, we present an enhancement to automatic summarization approaches of subroutines, by using the file context of the subroutines combined with the subroutines' code.  By ``file context'' we mean the other subroutines in the same file.  What we propose, in a nutshell, is to start with one of several existing approaches that models a subroutine's source code, then 1) model the signatures of all other subroutines in the same file also using recurrent nets, and 2) use an attention mechanism to learn associations between subroutines in the file context to words in the target subroutine's summary during training.  Our idea is novel because existing approaches generally only attend words in the output summary to words in the target subroutine.  Combined with other advancements that we will describe in this paper, our approach is able to learn much richer associations among words and produce better output summaries.  In our experimental section, we demonstrate that our approach improves existing baselines by providing orthogonal information to help the model provide better predictions for otherwise difficult to understand subroutines.

\vspace{-0.5cm}
\section{Problem, Significance, Scope}

The problem we target is called ``source code summarization'', a term coined by Haiduc~\emph{et al.}~\cite{haiduc2010use} around 2009 to refer to the process of writing short, natural language descriptions (``summaries'') of source code.  We, along with a majority of related work (see Section~\ref{sec:related}), focus on summarization of subroutines because subroutine summaries have a high impact on the quality of documentation~\cite{forward2002relevance} and because the problem of code summarization of subroutines is analogous to Machine Translation (MT) for which there is a large potential for cross-pollination of ideas with the natural language processing research area, as a number of new interdisciplinary workshops and NSF-funded meetings have highlighted~\cite{NL4SEAAAI:2018}.  To encourage this cross-pollination of ideas and to maximize the reproducibility of our work, we focus on Java methods and use a dataset recently prepared by LeClair~\emph{et al.} for NAACL'19~\cite{leclair2019recommendations}.  However, in principle our work may apply to many programming languages in which subroutines are organized into files.

Yet despite crossover with NLP, this work is firmly in the field of Software Engineering.  A very brief history of code summarization is that research efforts focused on manual rule-writing and heuristics until around 2016 when Neural Machine Translation was applied to source code and comments.  The first NMT applications treated code summarization as essentially an MT problem: source code was the ``source'' language while summary comments were the ``target'', compared to an MT setting when e.g. French sentences were a source and equivalent e.g. English sentences the target.  A trend since then has been to further define the problem in terms of software engineering domain-specific details.  For example, marking up the source language with data from the abstract syntax tree of the methods~\cite{alon2018code2seq, hu2018deep} or modeling the code as a graph rather than a sequence~\cite{allamanis2018learning}.  This paper moves further in that direction, by using file context to improve summarization.  Future work will likely continue this trend, with better results coming from better distinctions between source code summarization and machine translation.

It is important to recognize that we seek to \emph{enhance} existing approaches, rather than compete with them.  We present our approach as an augmentation to several existing baselines.  Also, beyond the solution we present, a scholarship objective of this paper is to accelerate the community's progress by demystifying key aspects of using neural networks for code summarization.  A frequent complaint about scientific literature and AI research specifically is that it is hard to reproduce and tends to treat the solutions as a black box~\cite{olden2002illuminating}.  We aim to push against that tendency.  Therefore, we dedicate several discussions in this paper to justifying decisions and explaining results in detail.

\section{Background \& Related Work}

This section discussion key background items including related work from software engineering and natural language processing.

\vspace{-0.2cm}
\subsection{Source Code Summarization}
\label{sec:related}

As mentioned above, source code summarization has a long history in software engineering literature, generally following the pattern of heuristic-based methods giving way to more recent data-driven methods.  Specifically, there are five very closely related data-driven papers on source code summarization which we cover in detail.

The first of the five closely-related papers is by Iyer~\emph{et al.}~\cite{iyer2016summarizing} published in 2016.  This work was one of the earliest to use a neural encoder-decoder architecture for code summarization.  The work set the foundation for significant advancements, but in retrospect was a fairly straightforward application of off-the-shelf NMT technology: C\# code was used as an input language and summary descriptions used as an output.  The paper made several changes to the input during preprocessing in an attempt to maximize performance by focusing on the important parts of the code.  In a sense this could be thought of as a bridge between pre-2016 heuristic-based approaches and later ``big data'' solutions: heuristics were used to select important words from code in preprocessing, then fed to an encoder-decoder system that, in it's overall structure, remained unchanged from NMT approaches for natural languages.

Hu~\emph{et al.}~\cite{hu2018deep} in 2018 proposed an improvement using the Abstract Syntax Tree (AST) of the function.  Their idea was that the AST should give more details about the code behavior than the words in the code alone, and therefore should lead to improved prediction of code summaries.  However the problem they encountered was that a vast majority of encoder-decoder models at the time relied on sequence input, while the AST is a tree.  Their solution was to design a Structure-Based Traversal (SBT) which is essentially an algorithm for flattening the AST into a sequence and using the components of the AST to annotate words from the code.

Next, LeClair~\emph{et al.}~\cite{leclair2019neural} at ICSE 2019 observed that the approach by Hu~\emph{et al.} blended two very different types of information (structure from the AST and language from identifier names etc.) in the same input sequence, while in other research areas such as image captioning~\cite{hossain2019comprehensive} it has been shown that better results are achieved when different input types are processed separately.  Therefore, they designed a model architecture that processes the word sequence and SBT/AST sequences in separate recurrent nets with separate attention mechanisms, and concatenates the results into a context vector just prior to prediction.

Meanwhile, Wan~\emph{et al.}~\cite{wan2018improving} report improvements with a hybrid AST+code attention approach.  They also show how to use reinforcement learning to improve performance by several percent.  While we do not dispute that the RL-based approach helps, we do not use it this paper because our goal is to show how file context adds orthogonal information to AST+code approaches.  The RL-based approach is more like an improved training procedure, as opposed to adding new information to the model.  Ultimately, we used LeClair~\emph{et al.}~\cite{leclair2019neural}'s approach as a baseline because it is simpler (to reduce experimental variables) and slightly more recent.

At around the same time, Alon~\emph{et al.}~\cite{alon2018code2seq} and Allamanis~\emph{et al.}~\cite{allamanis2018learning} noted that while the AST is a useful addition to the model for prediction, it is not optimal to flatten the AST into a sequence, since that forces the model to learn tree structure information from a sequence.  Yet these papers diverge substantially on the solution.  Alon~\emph{et al.} extract a series of paths in the AST and treat each path as a different sequence, while Allamanis~\emph{et al.} propose using a graph-based neural network to model the source code.  However, Allamanis~\emph{et al.} targeted generative models of the code itself.  In this sense, Allamanis~\emph{et al.}'s work is representative of a variety of neural models for code representation~\cite{yu2019neural, zhang2019novel, balog2017deepcoder}.

Beyond Allamanis~\emph{et al.}'s connection to code summarization with a recommendation of graph-based NNs, there is a diverse and growing body of work applying neural representations of code to several other problems such as commit message generation~\cite{jiang2017automatically}, pseudocode generation~\cite{oda2015learning}, and code search~\cite{gu2018deep}.  Due to space limitations we direct readers to peer-reviewed surveys by Chen~\emph{et al.}~\cite{chen2018best} and Allamanis~\emph{et al.}~\cite{allamanis2017survey} as well as an online running survey~\cite{ml4codewebsite}.  Song~\emph{et al.}~\cite{song2019survey} and Nazar~\emph{et al.}~\cite{nazar2016summarizing} describe code summarization in detail, including heuristic-based techniques~\cite{sridhara2010towards, sridhara2011automatically, mcburney2016automatic}.

\vspace{-0.4cm}
\subsection{Neural Encoder-Decoder Architecture}

The key supporting technology for nearly all published data-driven code summarization techniques is the neural encoder-decoder attention architecture, designed for Neural Machine Translation (NMT).  The origin of the architecture is Bahdanau~\emph{et al.}~\cite{bahdanau2014neural} in 2014.  While the encoder-decoder design existed, that paper introduced attention, which drastically increased performance and launched several new threads of research.  Since that time the number of papers using the attentional encoder-decoder design far exceeds what can be listed in one paper, with thousands of papers discussing overall improvements as well as adjustments for specific domains~\cite{chu2018survey}.  Nevertheless, this history has been chronicled in several surveys~\cite{young2018recent, pouyanfar2018survey, shrestha2019review}.

We include enough details in Section~\ref{sec:approach} to understand the basic encoder-decoder architecture and how our approach differs, but provide an overview here.  In general, what ``encoder-decoder'' means is that an ``encoder'' is provided input data representing a source sentence or document, while the decoder is provided samples of desired output for that input.  After sufficient training examples provided to the encoder and decoder, the model can often generate a reasonable output from the decoder given an input to the encoder.  In the most basic setup, the encoder and decoder are both recurrent neural networks: the encoder RNN's output state is given as the initial state of the decoder.  During training, the decoder learns to predict outputs one word at a time based on the encoder RNN's output state.  Advancements in the encoder-decoder model usually focus on the encoder because the encoder is what models the input, and more accurate input modeling is likely to lead to more accurate predictions of output.  For example, the encoder input RNN may be swapped for a graph-NN~\cite{allamanis2018learning,xu2018graph2seq,xu2018exploiting,chen2019reinforcement}.

The encoder-decoder design has found uses in a very wide variety of applications beyond its origin for translation.  Another application area relevant to this paper is document summarization, in which a paragraph or even several pages is condensed to one or two sentences.  Typical strategies include representing each sentence in the document with an RNN, and selecting words from these sentences in the decoder~\cite{nallapati2016abstractive}.  From a very high level this strategy is relevant to our work, in that we model each function in a file with an RNN, though numerous important differences will become apparent in the next section.

\section{Our Model}
\label{sec:approach}

In this section, we present our prediction model.  Note that we did not include certain optimizations such as pretrained word embeddings, multi-layer RNNs, etc.  These optimizations are tangential to the main objective of this paper: to evaluate file context as an improvement for neural code summarization.  Therefore we keep our model design simple to reduce experimental variables, while retaining the most important design features from related work.

\vspace{-0.2cm}
\subsection{Overview}
\label{sec:overview}

The model is based on the encoder-decoder architecture.  As in related work, for each function, we have a source code/text input as well as an AST input to the encoder, plus a summary input to the decoder.  But, we introduce a new input called ``file context'', which is the code/text from every other function in the same file.  In this subsection, we discuss an overview visualized below:

\begin{figure}[!h]
	\centering
	\vspace{-0.15cm}
	\includegraphics[width=0.4\textwidth]{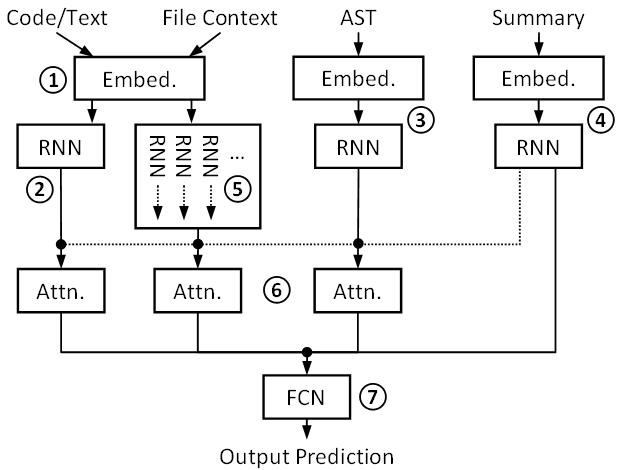}
	\vspace{-0.3cm}
	\label{fig:arch}
\end{figure}

The encoder can be thought of as modeling three types of input: code/text and the AST from a function, and the file context for that function.  Likewise the decoder models one type of input: the summary.  During training, a sample summary is provided to the decoder.  During inference, the decoder receives the predicted summary \emph{so far}, while the output prediction is the predicted next word in the summary (see the next subsection for training details).

\emph{Code/Text} We model code/text using a word embedding space and a recurrent network (area 2).  This is the same design that has been successful in several related papers.  The code/text is merely the sequence of tokens from the source code for the function.

\emph{AST} As discussed above, there are three ways related work models the AST: by flattening the tree to a sequence (Hu~\emph{et al.} and LeClair~\emph{et al.}), by using paths of the AST as separate sequences (Alon~\emph{et al.}), or by using a graph-NN (Allamanis~\emph{et al.}).  We built our model so that any one of these may be used, and provide implementations for each.  Later, our experiment will evaluate the effects of combining file context with each.  We mention a few caveats in this section, though.  First, we describe our implementation with the flattening technique shown by LeClair~\emph{et al.} (area 3) for simplicity and because, ultimately, we observed the highest BLEU scores in that configuration.  Second, for the AST encoder by paths, we were only able to use 100 paths, which is at the lowest limit recommended by Alon~\emph{et al.} (they found optimal setting was 200) -- when combining file context, we hit a known limit of CuDNN for large recurrent nets\footnote{https://github.com/tensorflow/tensorflow/issues/17009}.  Third, using graph-NNs expands the network size by tens of millions of parameters, leading to limitations on speed and memory usage; we chose two ``hops'' in the graph-NN as a compromise.


\emph{File Context} This input leads to the key novelty of our model.  The input itself is, essentially, the code/text data for every other function in the same file as the function we are trying to summarize.  The format of this input is an $n$x$m$ matrix in which $n$ is the number of other functions in the file and $m$ is the number of tokens in the function (technically, we pad or truncate functions at $m$ tokens so all input sequences are the same length, and select only the first $n$ functions from each file if more then $n$ are available).  We model the input first by using a trainable word embedding space (area 1).  Note that we share this embedding space with the code/text data -- there is not a separate vector space for words in the code/text input and words in the functions in the file context.  Note that this necessitates the use of the same vocabulary for both inputs, which has the effect of introducing more out-of-vocab words for the code/text input (since vocabulary is calculated as the $x$-most common words, and the file context distorts the word occurrence count in favor of words that occur in the file context).  Yet in pilot studies we found an improvement of over 20\% in BLEU score when sharing this space opposed to learning different embedding spaces, not to mention reduced model size and time to train.  For reference, we used a code/text vocab size of 75k, and about 15\% of this was displaced when recomputing the vocabulary with file context.

Next (area 5), we use one RNN per function in the file context.  This is similar to how function code/text is modeled (area 2), except that for file context we only use the final state of the RNN.  For function code/text (area 2), we output the RNN state for every position in the code/text sequence -- that way, we can compute attention for every position in the decoder (area 4) to every position in the code/text sequence (as described for NMT by Bahdanau~\emph{et al.} and implemented in a majority of code summarization papers).  In contrast, for file context, we build a two dimensional matrix in which every column is a vector representing a function in the file context (the vector is the final state of the RNN for that function).

As mentioned, we calculate attention (area 6) for each encoder input to the decoder (``summary'') input.  For code/text and AST sequences, we compute attention as mentioned in the previous paragraph.  However, for file context, we compute attention from each position in the summary to each function in the file.  A metaphor to NMT is that the attention mechanism was originally designed for building a dictionary between words in one language to words in another language, by learning to emphasize the positions in the encoder and decoder sequences that have the same meaning.  Essentially what this does is train the model to output a word in one language e.g. Hund when it sees e.g. dog in the input.  Applied to our model for file context, the model learns to output words in code summaries when it sees functions relevant to those words.  So for instance, if a function in the file context involves playing mp3 files, the model will be more likely to output words related to mp3s, music, audio, etc.  In our evaluation (see RQ$_2$), we explore evidence of how our model actually does behave as we envision.

To create an output prediction, we concatenate the attention-adjusted matrices from all three encoder inputs and the decoder input and use a fully-connected network (area 7) to learn how to combine the features from each input.  This part of the model is similar to most encoder-decoder networks, and ultimately outputs a prediction for the next word in the summary.

\vspace{-0.2cm}
\subsection{Training Procedure}

We train our model using the ``teacher forcing'' procedure described extensively in related work~\cite{doya2003recurrent, doya1992bifurcations}.  In short, this procedure involves learning to predict summaries one word at a time, while exposing the model only to the reference gold set summaries (as opposed to using the model's own predictions during training).  So for example, for every function, we train the model by providing the three encoder inputs, plus the decoder summary input one word at a time, e.g. a sample:

{\small
	\begin{verbatim}
	[ encoder inputs ] => play mp3 files
	\end{verbatim}
}

would become during training four separate samples:

{\small
	\begin{verbatim}
	[ encoder inputs ] => <st> + ( play )
	[ encoder inputs ] => <st> play + ( mp3 )
	[ encoder inputs ]	=> <st> play mp3 + ( files )
	[ encoder inputs ]	=> <st> play mp3 files + ( <et> )
	\end{verbatim}
}

where $<st>$ and $<et>$ are start and end sentence tokens (for readability, we do not show padding).  During training, the encoder would receive the encoder inputs, the decoder would receive the sequence so far (e.g. ``play mp3''), and sample output prediction would be the correct next word (e.g. ``files'') whether or not the model actually would have succeeded in making that prediction.

\vspace{-0.2cm}
\subsection{Corpus Preparation}
\label{sec:corpus}

We used the corpus provided by LeClair~\emph{et al.} in a NAACL'19 paper of recommendations for code summarization datasets~\cite{leclair2019recommendations}.  This corpus includes around 2m Java methods paired with summaries, already cleaned and split into training/validation/test sets according to a variety of recommendations.  That paper evaluated four different splits and determined minimal variations in reported results after cleaning.  For maximum reproducibility, we use ``split set 1'' from that paper.  We do not use datasets from other papers because we could not verify that they followed the dataset recommendations such as removing auto-generated code.

However, the corpus did not include file context (only code/text and AST) for each Java method.  Therefore, we obtained from Lopes~\emph{et al.}~\cite{Lopes+Bajracharya+Ossher+Baldi:2010} the raw data use by LeClair~\emph{et al.}, and created the $n$x$m$ file context matrices for each method (see paragraph 5 of Section~\ref{sec:overview}).  Note that each file context matrix did not include the function itself -- only the \emph{other} methods in the file.  We included these methods even if they did not have summaries of their own.  We filtered all comments and summaries from the file context so that it only included words from the code itself.  In practice it may be desirable to include these comments, but we felt that including them would create a possibility that the model could see a very similar or even identical summary during training, and we decided to avoid the possibility of introducing this bias.

Note that the size of $n$ and $m$ become important hyperparamters in our approach: $n$ controls the number of functions per file, and $m$ controls the number of tokens per function.  Ideally both numbers would be very high, but hardware and software limitations require us to cap them.  If $n$ is too high, many functions will be included, but they will all be very short (low $m$).  If $m$ is too high, we will only be able to model a few functions per file.  Ultimately we tested several values and found $n$=20, $m$=25 to provide a reasonable balance.  Note that the average number of methods per file in the corpus was about 8, and $n$=20 covers all functions in over 97\% of files.  Interestingly, performance plateaus after $m$=25.  The model appears to use the function signatures and first few tokens, but later parts of the function do not appear to be useful in the file context, at least with the type of RNNs we used in our implementation.

\vspace{-0.2cm}
\subsection{Model Details}
\label{sec:impl}

Inspired by successful examples set by previous work, we discuss our model details in the context of our implementation source code.  We built this model in a framework provided at ICSE'19~\cite{leclair2019neural}, and is readable via file {\small \texttt{atfilecont.py}} in our fork of that framework (with a few minor edits for readability, see details Section~\ref{sec:reproducibility}).

{\small \begin{verbatim}
tdat_input = Input(shape=(self.tdatlen,))
sdat_input = Input(shape=(self.n, self.m))
ast_input = Input(shape=(self.astlen,))
com_input = Input(shape=(self.comlen,))
\end{verbatim}}

First, above, are the input layers corresponding to the code/text, file context, and AST for the encoder, plus the summary comment for the decoder.  We discussed the $n$ and $m$ hyperparameters in the previous section.  For the others, we chose $tdatlen$=50 for the maximum number of tokens in the code/text sequence, $astlen$=100 for the max tokens in the flattened AST sequence, and $comlen$=13 for the max words in the output summary.  The parameters $tdatlen$ and $astlen$ are as recommended by LeClair~\emph{et al.} in their experiments and followup discussion with the authors, while $comlen$ is limited by the available corpus.

{\small \begin{verbatim}
tdel = Embedding(output_dim=100, input_dim=75000)
\end{verbatim}}

This line creates the code/text embedding space.  The space size is one 100-dimension vector for every 75k words.  We chose this size as a compromise between performance and memory usage.

{\small \begin{verbatim}
tde = tdel(tdat_input)
tenc = CuDNNGRU(256, return_state=True, return_sequences=True)
tencout, tstate_h = tenc(tde)
\end{verbatim}}

This section corresponds to area 2 of the overview figure.  Note that {\small \texttt{return\_sequences}} is enabled, meaning that the variable {\small \texttt{tencout}} will contain a matrix of size 50 x 256: one 256-length vector for each of the 50 positions in the code/text sequence.  Note that it is not typical for the RNN output dimensions (256 here) to exceed the word embedding vector length (100) for NMT applications, though we have repeatedly found a benefit in our pilot studies for code summarization tasks.

{\small \begin{verbatim}
de = Embedding(output_dim=100, input_dim=10908)(com_input)
dec = CuDNNGRU(256, return_sequences=True)
decout = dec(de, initial_state=tstate_h)
\end{verbatim}}

Next is the decoder (area 4 in overview figure).  The vocabulary size of 10908 is as provided in the corpus, though we note that it is less than the 44k reported by LeClair~\emph{et al.} in ICSE'19.  The reason for the change seems to be a greatly increased training speed at minimal performance penalty, but it does mean that the results are not directly comparable to other papers -- in our evaluation, we had to rerun the experiments with the new vocab size.

{\small \begin{verbatim}
ae = Embedding(output_dim=10, input_dim=100)(ast_input)
ae_enc = CuDNNGRU(256, return_sequences=True)
aeout = ae_enc(ae)
\end{verbatim}}

This is the AST input portion of the decoder (area 3).  Shown below is the SBT~\cite{hu2018deep} flat AST technique, but in our experiments we swap this section for other AST encoders (see Section~\ref{sec:overview}). 

{\small \begin{verbatim}
ast_attn = dot([decout, aeout], axes=[2, 2])
ast_attn = Activation('softmax')(ast_attn)
acontext = dot([ast_attn, aeout], axes=[2, 1])
tattn = dot([decout, tencout], axes=[2, 2])
tattn = Activation('softmax')(tattn)
tcontext = dot([tattn, tencout], axes=[2, 1])
\end{verbatim}}

The above is the attention mechanism for the code/text and AST inputs.  This part is traditional attention between each position in the decoder input to each position in the code/text and AST inputs.

{\small \begin{verbatim}
semb = TimeDistributed(tdel)
sde = semb(sdat_input)
\end{verbatim}}

These two lines begin our file context portion of the model (area 5).  Basically what happens is that one 25x100 matrix is generated for every function in the file context: that is, one 100-dimension vector for every one of the 25 (hyperparameter $m$) words in each function sequence.  The 100-dimension vectors are from the word embedding space shared with the code/text input (variable {\small \texttt{tdel}}).  There is one 25x100 matrix for each of the 20 (hyperparameter $n$) functions in the file, resulting in a 20x25x100 matrix as {\small \texttt{sde}}.

{\small \begin{verbatim}
senc = TimeDistributed(CuDNNGRU(256))
senc = senc(sde)
\end{verbatim}}

Next we create one RNN for each of the 20 functions.  Each RNN will receive a 25x100 matrix: 25 positions of 100-dimension word vectors.  Note that we also built a custom TimeDistributed layer in which we passed {\small \texttt{tstate\_h}} as the initial state for each RNN (as it is used for the decoder), but we noticed only minuscule performance differences and removed it for simplicity.  The size of {\small \texttt{senc}} is 20x256: one 256-length vector representing each of the 20 functions.

{\small \begin{verbatim}
sattn = dot([decout, senc], axes=[2, 2])
sattn = Activation('softmax')(sattn)
scontext = dot([sattn, senc], axes=[2, 1])
\end{verbatim}}

The attention mechanism for file context looks similar to the code/text and AST inputs, but has a very different meaning.  Variable {\small \texttt{sattn}} is the result of the dot product of {\small \texttt{decout}} and {\small \texttt{senc}}.  Consider the following multiplication (table format courtesy~\cite{leclair2019neural}):

\vspace{-0.4cm}
{\small
\begin{table}[h!]
	\begin{tabular}{p{2.1cm}lp{2.1cm}lp{2.1cm}}
		
		~~~~decout (axis 2) & &  ~~~~~senc (axis 2) & & ~~~~~~~~~sattn \\
		
		\begin{tabular}{p{0.00cm}p{0.00cm}p{0.00cm}p{0.00cm}p{0.00cm}}
			& 1        & 2        & ..        & 256   \\
			1  & \multicolumn{4}{l}{$v1 \longrightarrow$ } \\
			2  & \multicolumn{4}{l}{$v2 \longrightarrow$ } \\
			.. &          &          &           &       \\
			13 &          &          &           &           
		\end{tabular}
		& * &
		\begin{tabular}{p{0.00cm}p{0.00cm}p{0.00cm}p{0.00cm}p{0.00cm}}
			& 1        & 2        & ..        & 20   \\
			1   & $v3$                            & $v4$  &    &     \\
			2   & $\downarrow$ & $\downarrow$ &   &     \\
			..  &          &          &           &       \\
			256 &          &          &           &           
		\end{tabular}
		& = &
		\begin{tabular}{p{0.00cm}p{0.00cm}p{0.00cm}p{0.00cm}p{0.00cm}}
			& 1        & 2        & ..        & 20   \\
			1  & a		  & b		 & 			 & \\
			2  & c		  & d		 & 			 & \\
			.. &          &          &           &       \\
			13 &          &          &           &           
		\end{tabular}
		\\
		
	\end{tabular}
\end{table}
}
\vspace{-0.4cm}

In the above, vector $v1$ is the 256-dimension vector representing the first position in the decoder RNN.  The vector $v3$ is the 256-dimension vector representing the first function in the file context.  Value $a$ is a measure of similarity between those two vectors.  Vectors that are more similar will have a higher {\small \texttt{sattn}} matrix.

This similarity is important because we multiply {\small \texttt{sattn}} with {\small \texttt{senc}}:

{\small
\vspace{-0.4cm}
\begin{table}[h!]
	\begin{tabular}{p{2.1cm}lp{2.1cm}lp{2.1cm}}
		
		~~~~sattn (axis 2) & &  ~~~~senc (axis 1) & & ~~~~~scontext \\
		
		\begin{tabular}{p{0.00cm}p{0.00cm}p{0.00cm}p{0.00cm}p{0.00cm}}
			& 1        & 2        & ..        & 20   \\
			1  & \multicolumn{4}{l}{$v5 \longrightarrow$ } \\
			2  & \multicolumn{4}{l}{$v6 \longrightarrow$ } \\
			.. &          &          &           &       \\
			13 &          &          &           &           
		\end{tabular}
		& * &
		\begin{tabular}{p{0.00cm}p{0.00cm}p{0.00cm}p{0.00cm}p{0.00cm}}
			& 1        & 2        & ..        & 256   \\
			1   & $v7$                            & $v8$  &    &     \\
			2   & $\downarrow$ & $\downarrow$ &    &     \\
			..  &          &          &           &       \\
			20 &          &          &           &           
		\end{tabular}
		& = &
		\begin{tabular}{p{0.00cm}p{0.00cm}p{0.00cm}p{0.00cm}p{0.00cm}}
			& 1        & 2        & ..        & 256   \\
			1  & e		  & f		 & 			 & \\
			2  & g		  & h		 & 			 & \\
			.. &          &          &           &       \\
			13 &          &          &           &           
		\end{tabular}
		\\
		
	\end{tabular}
\end{table}
}
\vspace{-0.4cm}

Vector $v5$ is a list of similarities of position 1 in the summary to different functions in the file context.  E.g., element 3 of $v5$ is the similarity of position 1 in the summary to function 3 in the file.  In contrast, vector $v7$ contains all the first elements of the 256-dimension vectors representing different functions.  When each element of $v5$ is multiplied to the corresponding element in $v7$, the effect is to scale the element in $v7$ by the similarity represented in $v5$.  So if position 1 of the summary is very similar to function 3 of the file while not similar to other functions (i.e. element 3 of $v5$ is high while other elements in $v5$ are low), then that position will be retained while others attenuated.

Note above that we apply a softmax activation to {\small \texttt{sattn}}, so each of the vectors in that matrix (such as $v5$) will sum to one.  If, for example, element 3 in $v5$ is 0.90 and all others sum to 0.10, then it means that the product of the multiplication of $v5$ and $v7$ will include 90\% of the value of element 3, and only 10\% of the value of all other elements.  That product is the value $e$: it will be dominated by the values of $v7$ (first positions of the 256-dimension vectors for the functions) that are most similar to position 1 in the decoder.  Likewise, $f$ will be dominated by values of $v8$ (second positions of the function vectors) most similar to position 1 of the decoder, etc.  In this way, we create a context matrix {\small \texttt{scontext}} which includes one 256-dimension vector for each of the 13 decoder positions -- each of these vectors represents the functions most relevant to the word in that position of the summary.

{\small \begin{verbatim}
context = concatenate([scontext, tcontext, acontext, decout])
squash = TimeDistributed(Dense(256, activ="relu"))(context)
\end{verbatim}}

We concatenate the context matrices from each attention mechanism (and the decoder) into a single context matrix along axis 1, to create a matrix of size 13x1024 (since each smaller context matrix is 13x256).  We then use one fully-connected layer of size 256 to squash the 1024 matrix into a lower dimension.  This is common in encoder-decoder architectures in order to prevent overfitting, though it serves an additional purpose in our model of helping the model learn how to combine the information from each of the three encoder inputs.

{\small \begin{verbatim}
squash = Flatten()(squash)
out = Dense(10908, activation="softmax")(squash)
\end{verbatim}}

Finally, we flatten the 13x256 ``squashed'' context matrix into a single 3328-dimension vector so that we can connect it to a fully-connected output layer.  The output layer size is the vocabulary size, and the argmax of this output layer is the index of the predicted next word in the vocabulary.  Note that a large number of parameters occur between the distributed dense layer and the output layer (3328 to 10908 elements is over 36m parameters).  Significant time could be gained by reducing the vocab size or the size of the ``squash'' layer (the 1024 could be squashed to, say, 128 instead of 256), but at unknown cost to performance.  In the end, we keep these values consistent across all approaches in our experiments, to ensure an ``apples to apples'' comparison, even if optimizations could be made depending on user circumstances.

\vspace{-0.2cm}
\subsection{Hardware/Software Details}

Our hardware included two workstations with Xeon E1530v4 CPUs, 128gb RAM, and dual Quadro P5000 GPUs.  Software platform included Ubuntu 18.04.2, Tensorflow 1.14, CUDA 10.0, and CuDNN 7.  The implementation above is for Keras 2.2.4 in Python 3.6.
\section{Experiment Design}
\label{sec:experiment}

This section discusses the design of our experiment including research questions, methodology, and other conditions.

\vspace{-0.2cm}
\subsection{Research Questions}

Our research objective is to evaluate whether the our proposed mechanism for including file context in code summarization improves strong, recent baselines.  In particular, we aim to establish whether any improvement can be attributed to the file context, so we aim to reduce the number and effect of other factors.  We ask the following Research Questions (RQs):

\begin{description}
	\item[RQ$_{1}$] What is the performance of our proposed approach compared to recent baselines in terms of quantitative metrics in a standardized dataset?
	
	\vspace{0.05cm}
	
	\item[RQ$_{2}$] To what extent can differences in performance be attributed to the inclusion of file context?
\end{description}

The rationale for RQ$_1$ is straightforward: to compare our approach to existing approaches.  The scope of this question is to compare our work to relevant data-driven technologies, as opposed to heuristic-based/template solutions.  Generally speaking, it would not be a ``fair'' comparison for heuristic-based techniques because a heuristic could produce a valid summary that would not be anything like the reference solution.  The way to evaluate a heuristic approach is with a human study, but the scale of the dataset (e.g. over 90k samples in the test set) is much too large.  Therefore we follow precedent set in both the SE and NLP research areas, and use quantitative metrics to evaluate performance in broad terms over the whole test set.

However, there are many factors that can affect performance between one data-driven approach and another.  For example, the choice of exactly which type of recurrent unit to use (e.g. LSTM vs GRU, or uni-directional vs bi-directional) or the number of units in a hidden layer.  We control as many of these factors as possible by configuring the approaches in as similar a way as reasonable (see Section~\ref{sec:baselines}), but there is always a question as to whether a proposed variable is actually the dominant one.  Therefore, we ask RQ$_2$ to study how file context contributed to predictions in the model.

\vspace{-0.2cm}
\subsection{Methodology}

To answer RQ$_1$, we follow the methodology established by many papers on code summarization in both SE and machine translation in NLP.  We obtain a standard dataset (see Section~\ref{sec:corpus}), then use several baselines plus our approach to create predictions on the dataset, then compute quantitative metrics for the output.  For our approach, we trained for ten epochs and selected the model that had the best accuracy score on the validation set.  Unless otherwise stated in Section~\ref{sec:baselines}, this is also how we trained the baselines.

The quantitative metrics we use are BLEU~\cite{Papineni:2002:BMA:1073083.1073135} and ROUGE~\cite{lin2004rouge}.  These two metrics have various advantages and disadvantages, but one or another form the foundation of nearly all experiments on neural-based translation or summarization.  Both are basically measures of similarity between a predicted sentence and a reference.  BLEU creates a score based on matches of unigrams, bigrams, 3-grams, and 4-grams of words in the sentences.  ROUGE encompasses a variety of metrics such as gram matches and subsequences.  We report BLEU (1-4) as well as ROUGE-LCS (longest common subsequence) to provide good coverage of metrics without redundancy.  In cases when the reference is only three words long (the dataset has a minimum summary length of three words), we calculate only BLEU 1-3 and do not include BLEU 4 in the total BLEU score for that sentence, because otherwise the total BLEU score would be zero even for correct three-word predictions.

We answer RQ$_2$ in three stages.  First, we provide an overview comparison of predictions by different models to determine whether the inputs (code/text, AST, file context) give orthogonal results and estimate the proportion of predictions may be helped by file context.  Second, we extract specific examples to illustrate how the approach works in practice.  Whenever possible, we support our claims with quantitative evidence, to minimize the potential of bias.  One key piece of evidence in the examples is from the attention layer to file context (area 6 in the overview figure, {\small \texttt{scontext}} in Section~\ref{sec:impl}).  The attention layer shows which of the functions in the file context are contributing the most to the prediction.  By showing that a prediction is improved when when a particular function is attended, we can demonstrate that the file context is the most likely explanation for the improvement.  Finally, we explore evidence that it is prevalent for file context to improve predictions in a similar way as the example, and perform an ablation study in which we train and test the model using only AST and file context.


\vspace{-0.2cm}
\subsection{Baselines}
\label{sec:baselines}

We use five baselines.  We chose these baselines because 1) they are recent, 2) they represent both code-only and AST+code neural approaches that our approach is capable of enhancing, and 3) they had reproducibility packages.  Space limitations prevent us from listing all relevant details, so we provide complete implementations in our online appendix (Section~\ref{sec:reproducibility}).

{\small \textbf{\texttt{attendgru}}}  This baseline represents a ``no frills'' attentional encoder-decoder model using unidirectional GRUs.  It represents approaches such as Iyer~\emph{et al.}~\cite{iyer2016summarizing}, but this implementation was published by LeClair~\emph{et al.}~\cite{leclair2019recommendations}.  We configure {\small \texttt{attendgru}} with identical hyperparameters to our approach whenever they overlap (e.g. word embedding vector size).

{\small \textbf{\texttt{ast-attendgru}}}  This is the approach LeClair~\emph{et al.}~\cite{leclair2019neural} propose, using their recommended hyperparameters.  This approach is an enhancement of SBT~\cite{hu2018deep}, so we only compare against this approach.

{\small \textbf{\texttt{transformer}}}  Vaswani~\emph{et al.}~\cite{vaswani2017attention} proposed an attention-only (no recurrent steps) machine translation model in 2017.  It was received in the NLP community with significant fanfare so, given the success of the model for NMT, we evaluate it as a baseline.

{\small \textbf{\texttt{graph2seq}}}  Allamanis~\emph{et al.}~\cite{allamanis2018learning} proposed using a graph-NN to model source code, but applied it to a code generation task in their implementation.  To reproduce the idea as a baseline for code summarization, we use a graph-NN-based text generation system proposed by Xu~\emph{et al.}~\cite{xu2018graph2seq}.  We use the nodes and edges of the AST as input, with all other configuration parameters as recommended for NMT tasks by Xu~\emph{et al.}.

{\small \textbf{\texttt{code2seq}}}  Alon~\emph{et al.}~\cite{alon2018code2seq}, described in Section~\ref{sec:related}, is a recent code summarization approach based on AST paths that is reported to have good results on a C\# dataset.  We reimplemented the approach from Section~3.2 of their paper, with their online implementation as a guide.  Note we did \emph{not} use their implementation verbatim.  They had many other architecture variations, preprocessing, etc., that we had to remove to control experimental variables.  Otherwise, it would not have been possible to know whether performance differences were due to file context or these other factors.   




\section{Experiment Results}

\begin{table*}[]
	\caption{Performance summary in terms of BLEU aggregate (A) and 1-4, plus ROUGE Longest Common Subsequence precision, recall, and F1 measure.  Except for Transformer, all implementations were identical except for different AST and file context encoders.  For example, the \texttt{ast-attendgru} model included the RNN-based code/text encoder and the flattened AST encoder, and achieved 18.69 aggregate BLEU score. Key takeaway is that the ``FC'' file context models obtained higher performance than their default ``non FC'' counterparts.  BLEU-A is bold because it usually serves as the most important performance metric.}
	\label{tab:rq1}
	\vspace{-0.2cm}
	\begin{tabular}{l|cc|ccc|c|lllll|llll}
		& \multicolumn{2}{l|}{function text}                    & \multicolumn{3}{c|}{AST}                                                            & \multirow{2}{*}{\begin{tabular}[c]{@{}c@{}}file\\ context\end{tabular}} & \multicolumn{5}{c|}{BLEU}                                                                                              & \multicolumn{3}{c}{ROUGE-LCS}                                         &  \\ \cline{2-6}
		& \multicolumn{1}{l|}{rnn} & \multicolumn{1}{l|}{xform} & \multicolumn{1}{l|}{flat} & \multicolumn{1}{l|}{graph} & \multicolumn{1}{l|}{paths} &                                                                         & \multicolumn{1}{c}{A} & \multicolumn{1}{c}{1} & \multicolumn{1}{c}{2} & \multicolumn{1}{c}{3} & \multicolumn{1}{c|}{4} & \multicolumn{1}{c}{P} & \multicolumn{1}{c}{R} & \multicolumn{1}{c}{F} &  \\
		transformer      & \multicolumn{1}{l}{}     & x                          & \multicolumn{1}{l}{}      & \multicolumn{1}{l}{}       & \multicolumn{1}{l|}{}      & \multicolumn{1}{l|}{}                                                   & \textbf{5.43}         & 22.46                 & 7.78                  & 2.73                  & 1.82                   & 28.72                 & 28.26                 & 27.62                 &  \\
		attendgru        & x                        &                            &                           &                            &                            &                                                                         & \textbf{18.22}        & 37.69                 & 20.89                 & 13.70                 & 10.22                  & 54.69                 & 47.42                 & 49.01                 &  \\
		ast-attendgru    & x                        &                            & x                         &                            &                            &                                                                         & \textbf{18.69}        & 37.13                 & 21.11                 & 14.27                 & 10.90                  & 56.60                 & 47.42                 & 49.75                 &  \\
		graph2seq        & x                        &                            &                           & x                          &                            &                                                                         & \textbf{18.61}        & 37.56                 & 21.27                 & 14.13                 & 10.63                  & 55.32                 & 47.45                 & 49.29                 &  \\
		code2seq         & x                        &                            &                           &                            & x                          &                                                                         & \textbf{18.84}        & 37.49                 & 21.36                 & 14.37                 & 10.95                  & 56.15                 & 47.55                 & 49.69                 &  \\
		attendgru+FC     & x                        &                            &                           &                            &                            & x                                                                       & \textbf{19.36}        & 37.40                 & 21.58                 & 14.94                 & 11.65                  & 57.11                 & 47.74                 & 50.19                 &  \\
		ast-attendgru+FC & x                        &                            & x                         &                            &                            & x                                                                       & \textbf{19.95}        & 38.39                 & 22.18                 & 15.41                 & 12.06                  & 56.19                 & 48.04                 & 50.03                 &  \\
		graph2seq+FC     & x                        &                            &                           & x                          &                            & x                                                                       & \textbf{18.88}        & 37.12                 & 21.21                 & 14.46                 & 11.17                  & 55.22                 & 47.03                 & 49.07                 &  \\
		code2seq+FC      & x                        &                            &                           &                            & x                          & x                                                                       & \textbf{19.11}        & 37.17                 & 21.39                 & 14.69                 & 11.41                  & 56.77                 & 47.40                 & 49.87                 & 
	\end{tabular}
	\vspace{-0.2cm}
\end{table*}

\begin{figure}[b!]
	\vspace{-0.3cm}
	\centering
	\includegraphics[width=7cm]{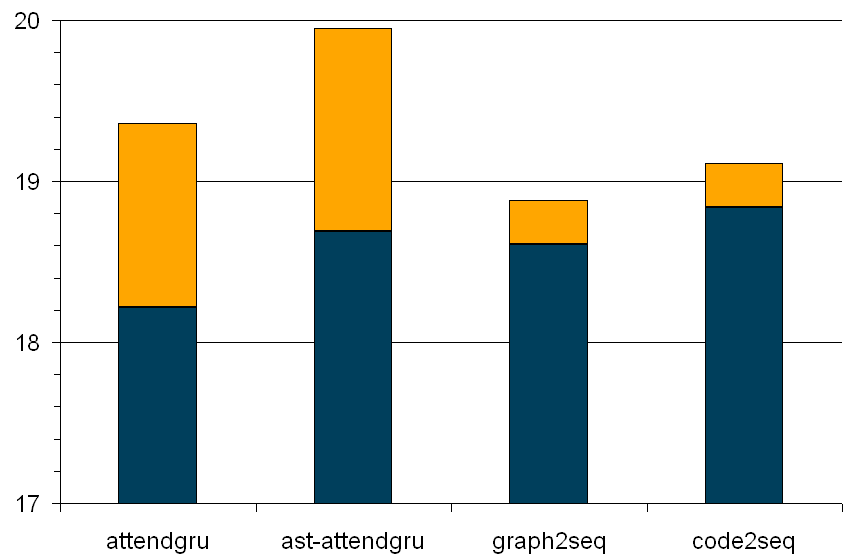}
	\vspace{-0.2cm}
	\caption{Comparison of baselines to baselines including our file context encoder.  Dark blue indicates baseline aggregate BLEU score (compare to Table~\ref{tab:rq1} column BLEU-A).  Orange indicates increase in BLEU score for identical model with file context encoder added.  Note y-axis starts at 17 BLEU.}
	\label{fig:rq1}
\end{figure}

We present our experimental findings and answer our research questions in this section.  Recall that our research objective is to evaluate the effects of adding our file context encoding, and to this end we present a mixture of high level quantitative data and specific examples and qualitative explanations.  Recall that we do not view our results as in competition with the baselines, but instead as a complementary attempt at improvement.

\vspace{-0.1cm}
\subsection{RQ$_1$: Quantitative Measures}

Our key finding in answering RQ$_1$ is that, in terms of quantitative measures over the whole test set, adding our file context encoder increased performance in nearly all cases.  Figure~\ref{fig:rq1} showcases the difference when compared with aggregate BLEU score (column BLEU-A of Table~\ref{tab:rq1}).  The baselines {\small \texttt{attendgru}} and {\small \texttt{ast-attendgru}} improved by more than one full BLEU point, while {\small \texttt{graph2seq}} and {\small \texttt{code2seq}} improved by around 0.3 BLEU.  One possible explanation for the greater increase for {\small \texttt{attendgru}} and {\small \texttt{ast-attendgru}} is that the path- and graph-based AST encoder models have many millions more parameters than the flattened AST approach (not to mention, when there is no AST encoder), and the model may have difficulty retaining some of the details in these larger encoders in the ``squash'' layer of the model (area 7 of overview figure in Section~\ref{sec:overview} and paragraph 13 of Section~\ref{sec:impl}).  Recall that an overriding objective in our experimental setup is to reduce variables to create an ``apples to apples'' comparison.  For that reason, we used fully-connected networks of 256 dimensions for all approaches in the squash layer.  The result could be that the model is able to learn more details about the flattened AST encoder in the squash layer simply because of there are many fewer parameters in that encoder, while we are in effect asking the model to remember much more information in that layer in the path- and graph-based encoders.  Our recommendation for future work is to designate the size of the squash layer as a hyperparameter for tuning, perhaps with larger settings for larger encoders.  We do \emph{not} recommend concluding from this experiment alone that a flat AST encoder is the best design.  Instead, we confine ourselves to the conclusion that adding our file context encoder to the baselines improves the performance of those baselines.

An exception to the overall observation of improved performance with the file context encoder is that the {\small \texttt{graph2seq}} model obtain a slightly lower ROUGE-LCS F1 score, despite a higher aggregate BLEU score, when we added the file context encoder.  The way to interpret ROUGE-LCS is that precision and recall are calculated only for words in a predicted sentence that appear in sequence when compared to a reference sentence.  For example, consider prediction ``converts the file from mp3 to wav'' and reference ``converts the file from wav to mp3''.  The summaries have very different meanings, but the BLEU1-3 scores will be fairly high and will inflate the aggregate BLEU score.  In contrast, ROUGE-LCS will result in precision and recall scores of 57\% (4/7 for longest common subsequence divided by sentence lengths, see Section 3 of Lin~\emph{et al.}~\cite{lin2004rouge} for formulae) -- the result is a score much more based on proper word order instead of predicting correct words anywhere in the sentence.  So, it appears that the predictions from {\small \texttt{graph2seq}} have slightly longer common subsequences with the references, without the file context encoder.

We made several related but tangential observations.  We found that the {\small \texttt{transformer}} model performed quite poorly on this software dataset, despite reports of excellent performance at low cost on natural language translation~\cite{vaswani2017attention, young2018recent}.  While it is tempting to draw sweeping conclusions from this finding, we mention it only to recommend caution in applying NMT solutions to this problem.  Evidence is accumulating in related literature that code summarization is not merely an application of off-the-shelf technology borrowed from the NLP domain~\cite{hellendoorn2017deep}.  
Another tangential observation is that we verify conclusions by related work, namely Alon~\emph{et al.}~\cite{alon2018code2seq} that a path-based AST encoder outperforms most alternatives ({\small \texttt{code2seq}} was the highest performer without file context), and Hu~\emph{et al.}~\cite{hu2018deep} and LeClair~\emph{et al.}~\cite{leclair2019neural} that a flat AST encoder design outperforms traditional seq2seq encoder-decoder designs (i.e. {\small \texttt{attendgru}}).  Finally, we note that the AST-based models have broadly similar performance (18.61 - 18.84 BLEU), while the contribution of file context varies much more (18.88 - 19.95 BLEU).

\subsection{RQ$_2$: Effects of File Context}
\label{sec:rq2}

\begin{figure}[b!]
	\vspace{-0.4cm}
	\centering
	\includegraphics[width=8.4cm]{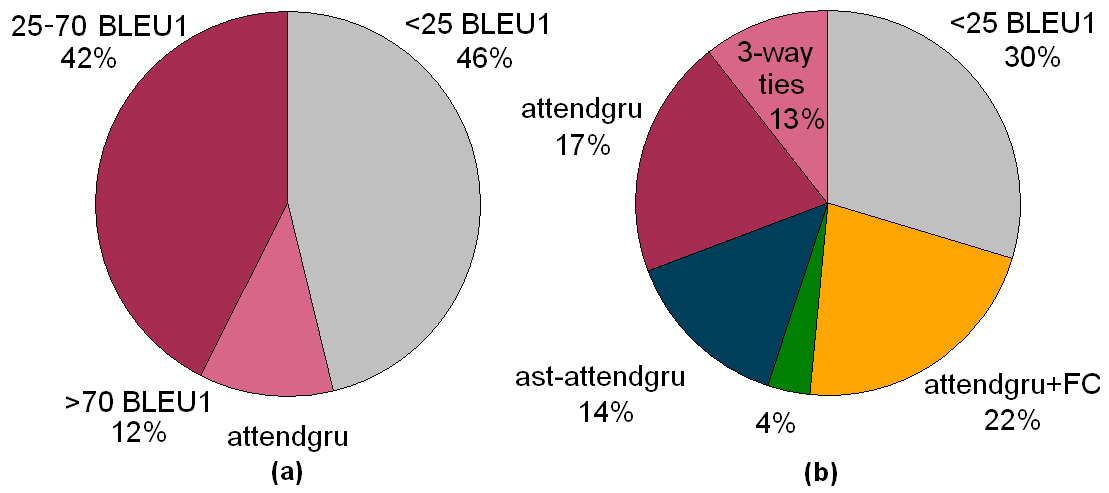}
	\vspace{-0.4cm}
	\caption{Comparison of predictions from models with code/text only ({\small \texttt{attendgru}}), code/text+AST ({\small \texttt{ast-attendgru}}), and code/text+FC ({\small \texttt{attendgru+FC}}).  Details in Section~\ref{sec:rq2}.}  
	\label{fig:rq2ab}
\end{figure}

We explore the effects of file context in three ways: First we offer a bird's eye view of results.  Second we present specific examples demonstrating how the model works.  Third we provide evidence that the examples are representative of the model's behavior.

\textbf{Overview} Consider Figure~\ref{fig:rq2ab}(a) which shows a breakdown of BLEU1 scores for predictions by {\small \texttt{attendgru}}.  (We show BLEU 1 scores for simplicity of comparison, BLEU 1 is equivalent to unigram precision.)  There is a significant portion in which {\small \texttt{attendgru}} performs quite well, with BLEU1 scores sometimes even nearing 100.  This is not surprising since in some cases, the source code of the function ``gives away'' the summary e.g. in functions with names like {\small \texttt{convertMp3ToWav}}.  Yet for 46\% of predictions, BLEU1 score is less than 25, meaning that not even 25\% of the words in the predictions are correct.  This is also not surprising, since in many cases the code has almost no clues as to what words should be in the summary.

Next consider Figure~\ref{fig:rq2ab}(b).  This chart shows the percent of functions in which the predictions had the highest BLEU1 score (of predictions >25 BLEU1) for three different models: {\small \texttt{attendgru}} relies only on code/text, {\small \texttt{ast-attendgru}} is the same model but with access to AST information, and {\small \texttt{attendgru+FC}} is the same model but with access to file context (but no AST).  What we observe is that there is a subset of functions for which each model seems to perform best -- it is not as through the models provide uniformly better results on all functions.  Related work e.g. LeClair~\emph{et al.}~\cite{leclair2019neural} showed how AST-based models can improve predictions for functions in which the structure contains clues about the function's behavior, even if the source code contains no useful words.  We make a similar observation for file context.  The {\small \texttt{attendgru+FC}} model performs best for a subset of around 22\% of functions (and ties with {\small \texttt{ast-attendgru}} for 4\% of functions).  When combined, a major portion of the improvement comes from reducing the number of low quality predictions (visible in a reduced number of <25 BLEU1 scores) in addition to improving many of the predictions of the baselines.  The result is the overall improvement in BLEU scores reported for RQ$_1$ for models that combine many types of input data such as {\small \texttt{ast-attendgru+FC}}.

\begin{figure}[b!]
	\setcounter{figure}{0}
	\renewcommand{\figurename}{Example}
	\vspace{-0.3cm} 
	{
		
		\begin{verbatim}
  public void setIntermediate(String intermediate) {
         this.intermediate = intermediate;    }
		\end{verbatim}
	}
	\vspace{0.2cm}
	{\small	
		\begin{tabular}{lm{5.5cm}} 
			\emph{reference} & sets the intermediate value for this flight  	\\ \hline
			attendgru        & sets the intermediate value for this <UNK>   	\\ 
			ast-attendgru    & sets the intermediate value for this <UNK>		\\ 
			ast-attendgru+FC & sets the intermediate value for this flight 		\\ 

		\end{tabular}
	}

	{\small
		\vspace{0.3cm}
		

	\begin{tabular}{lp{0.8cm}p{1mm}p{0.75mm}p{0.1mm}p{1.75mm}p{0.75mm}p{0.75mm}p{0.75mm}p{0.75mm}p{0.75mm}p{0.75mm}p{0.75mm}p{0.75mm}p{0.75mm}}
	\multicolumn{2}{l}{<st>}	& 1  & \multicolumn{12}{l}{\multirow{13}{*}{\includegraphics[width=5cm]{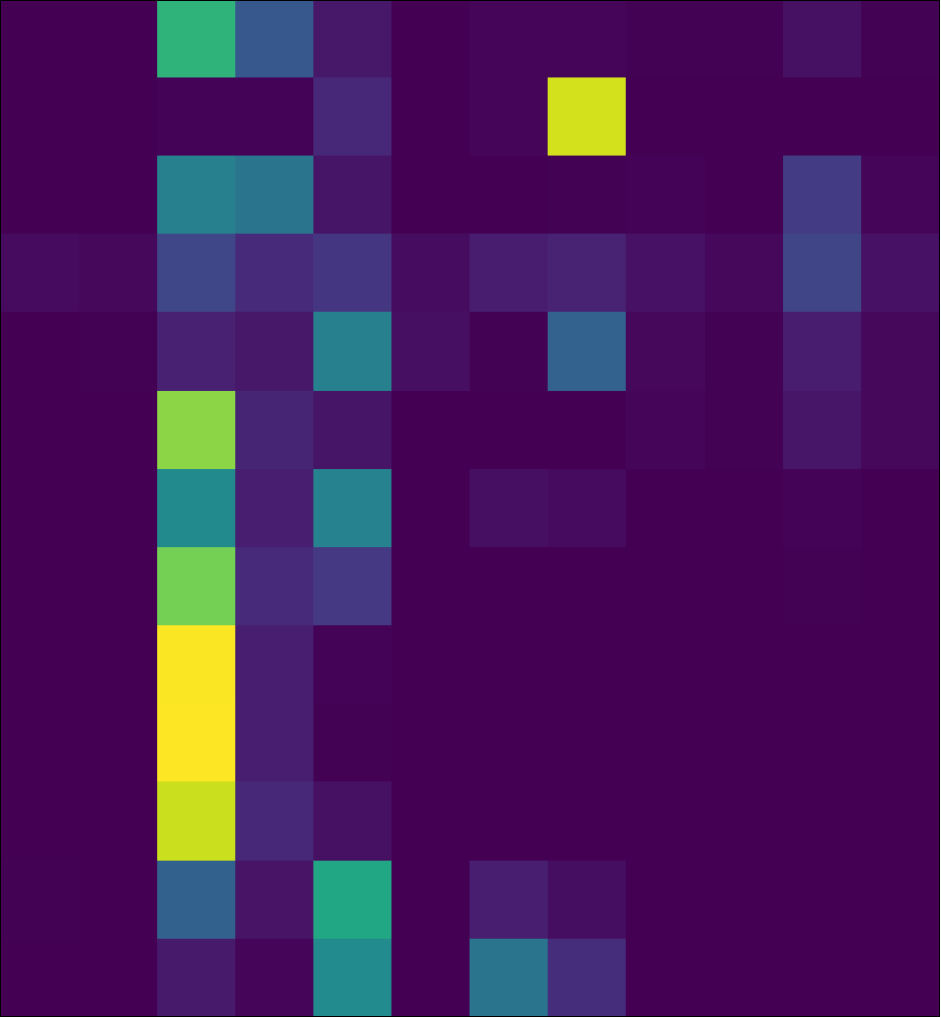}}}          \\[2pt]
	\multicolumn{2}{l}{sets}	& 2  & \multicolumn{12}{l}{}                            \\[2pt]
	\multicolumn{2}{l}{the}	& 3  & \multicolumn{12}{l}{}                            \\[2pt]
	\multicolumn{2}{l}{intermediate}	& 4  & \multicolumn{12}{l}{}                            \\[2pt]
	\multicolumn{2}{l}{value}	& 5  & \multicolumn{12}{l}{}                            \\[2pt]
	\multicolumn{2}{l}{for}	& 6  & \multicolumn{12}{l}{}                            \\[2pt]
	\multicolumn{2}{l}{this}& 7  & \multicolumn{12}{l}{}                            \\[2pt]
	\multirow{6}{*}[17pt]{\Bigg\downarrow} & \multirow{7}{*}[23pt]{\makecell{\hspace{-0.3cm}\emph{predicting} \\ \hspace{-0.3cm}\emph{next word}}} 	  & 8  & \multicolumn{12}{l}{}                            \\[2pt]
	&	& 9  & \multicolumn{12}{l}{}                            \\[2pt]
	&	& 10 & \multicolumn{12}{l}{}                            \\[2pt]
	&	& 11 & \multicolumn{12}{l}{}                            \\[2pt]
	&	& 12 & \multicolumn{12}{l}{}                            \\[2pt]
	&	& 13 & \multicolumn{12}{l}{}                            \\[3pt]
	&	&    & 1 & 2 & \circled{3} & 4 & 5 & 6 & 7 & 8 & 9 & 10 & 11 & 12
	\end{tabular}
		
		\vspace{0.2cm}
		
	\begin{tabular}{ll}
		1  & public void set airline name java lang string airline name this ...    \\
		2  & public void set destination java lang string destination this ...         \\
		\hspace{-0.08cm}\circled{3}  & public long get flight id return flight id                                                    \\
		4  & public void set flight id long flight id this flight id flight id                             \\
		5  & public void set flight number java lang string flight number this ... \\
		6  & public void set intermediate java lang string intermediate ...     \\
		7  & public void set intermediate arrival time java lang string ...                       \\
		8  & public void set intermediate departure time java lang string ...                     \\
		9  & public int get num available seats return num available seats                                 \\
		10 & public void set num available seats int num available seats                                   \\
		11 & public int get num seats return num seats                                                     \\
		12 & public void set num seats int num seats                                                      
	\end{tabular}

	\vspace{-0.3cm}

\caption{(upper) Source code, summaries, and predictions.  (mid) activation map of the attention matrix in {\small \texttt{ast-attendgru+FC}} from summary to files just prior to predicting position 8, the word ``flight''.  The x-axis is the position in the summary vector.  The y-axis is the function in the file.  (lower) Finally, inputs to the file context matrix.}

}
\end{figure}

\textbf{Examples} Below are two examples showing how using file context improves predictions.  We chose these examples based on explanatory power: they are short methods in which {\small \texttt{ast-attendgru}} and {\small \texttt{ast-attendgru+FC}} differed by one word (there are 109 such examples out of 6945 where {\small \texttt{ast-attendgru+FC}} outperformed {\small \texttt{ast-atte- ndgru}} over the 90908 methods in the test set).

In Example 1, it is obvious from the code that the method is just a setter, but there is no hope to understand what the value means even with the AST, and the baselines output an unknown token.  But, the file context reveals several keywords e.g. flight, airline, departure that serve as clues.  The attention matrix shows that the model found these clues (the image shows a heatmap of the values in {\small \texttt{sattn}} in Section~\ref{sec:impl} just prior to predicting the last word).  High activation is visible connecting the later positions in the output prediction to function 3, which contains the word ``flight.''  The effect of high activation on this function is that the context vector (which is a concatenation of the attention matrices) will be much closer in vector space to the words related to flights, airlines, etc. in the word embedding, than it would be with only the code/text and AST.  This makes it much easier for the model to predict the correct word.

\begin{figure}[b!]
	\setcounter{figure}{1}
	\renewcommand{\figurename}{Example}
	\vspace{-0.3cm} 
	{
		
		\begin{verbatim}
  public String toString() {
         if (throwable != null) 
               return super.toString() 
                   + System.getProperty("line.separator") 
                + throwable.toString();
         return super.toString();    }
		\end{verbatim}
	}
	\vspace{0.2cm}
	{\small	
		\begin{tabular}{lm{5.5cm}} 
			\emph{reference} & returns a string representation of this exception  	\\ \hline
			attendgru        & returns a string representation of this object   	\\ 
			ast-attendgru    & returns a string representation of this object		\\ 
			ast-attendgru+FC & returns a string representation of this exception 		\\ 
			
		\end{tabular}
	}
	
	{\small
		\vspace{0.3cm}
		

		\begin{tabular}{lp{0.8cm}p{1mm}p{0.75mm}p{0.1mm}p{0.75mm}p{0.75mm}p{1.75mm}p{0.75mm}p{0.75mm}p{0.75mm}p{0.75mm}p{0.75mm}p{0.75mm}p{0.75mm}}
			\multicolumn{2}{l}{<st>}	& 1  & \multicolumn{12}{l}{\multirow{13}{*}{\includegraphics[width=2.5cm]{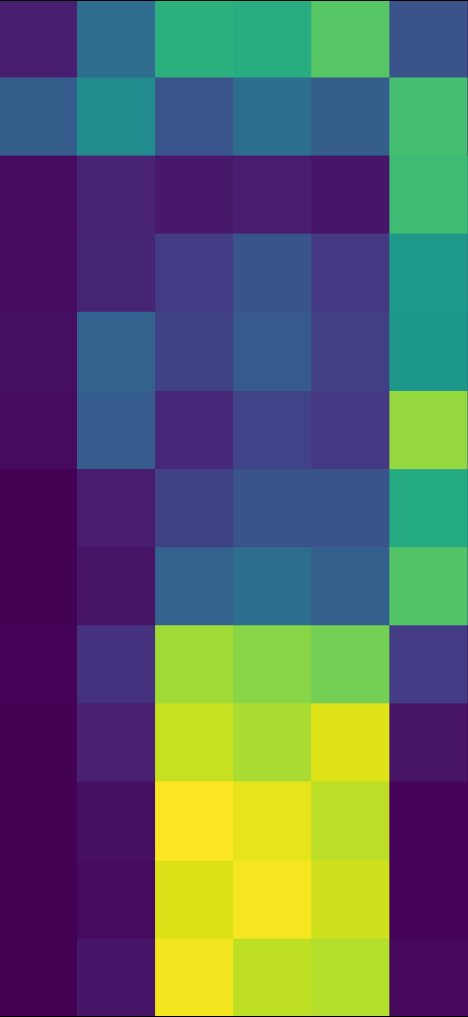}}}          \\[2pt]
			\multicolumn{2}{l}{returns}	& 2  & \multicolumn{12}{l}{}                            \\[2pt]
			\multicolumn{2}{l}{a}	& 3  & \multicolumn{12}{l}{}                            \\[2pt]
			\multicolumn{2}{l}{string}	& 4  & \multicolumn{12}{l}{}                            \\[2pt]
			\multicolumn{2}{l}{representation}	& 5  & \multicolumn{12}{l}{}                            \\[2pt]
			\multicolumn{2}{l}{of}	& 6  & \multicolumn{12}{l}{}                            \\[2pt]
			\multicolumn{2}{l}{this}& 7  & \multicolumn{12}{l}{}                            \\[2pt]
			\multirow{6}{*}[17pt]{\Bigg\downarrow} & \multirow{7}{*}[23pt]{\makecell{\hspace{-0.3cm}\emph{predicting} \\ \hspace{-0.3cm}\emph{next word}}} 	  & 8  & \multicolumn{12}{l}{}                            \\[2pt]
			&	& 9  & \multicolumn{12}{l}{}                            \\[2pt]
			&	& 10 & \multicolumn{12}{l}{}                            \\[2pt]
			&	& 11 & \multicolumn{12}{l}{}                            \\[2pt]
			&	& 12 & \multicolumn{12}{l}{}                            \\[2pt]
			&	& 13 & \multicolumn{12}{l}{}                            \\[3pt]
			&	&    & 1 & 2 & \circled{3} & \circled{4} & \circled{5} & 6 &   &   &   &   &   &  
		\end{tabular}
		
		\vspace{0.2cm}
		
		\begin{tabular}{ll}
			1  & public void set throwable throwable a throwable this throwable ...    \\
			2  & public throwable get throwable return throwable        \\
			\hspace{-0.08cm}\circled{3}  & public void print stack trace if throwable null throwable print ...                                          \\
			\hspace{-0.08cm}\circled{4}  & public void print stack trace print stream s if throwable null ...                             \\
			\hspace{-0.08cm}\circled{5}  & public void print stack trace print writer s if throwable null ...     \\
			6  & public string to string if throwable null return super to string ...  \\                                                    
		\end{tabular}
		
		\vspace{-0.3cm}
		
		\caption{Setup similar to Example 1.  Note significant attention to functions 3, 4, and 5 in file context, which contain the term ``stack trace.''}
		
	}
\end{figure}

Example 2 is similar, except that the word that is different in the predictions (``exception'' vs ``object'') is not directly in the file context.  Once again the source code gives few clues except that it returns a string representation of something.  The baselines give the term ``object'' which is a reasonable guess but not specific.  The attention matrix shows high activation on the three functions which contain the term ``stack trace'', which is nearby to ``exception'' in the word embedding (technically ``stack trace'' is two words, but since the function representation is an RNN, the final state will contain information from both words).

\textbf{Prevalence} The examples above show how file context \emph{can} improve predictions in specific cases, but we chose these examples based on explanatory power and not necessarily prevalence.  In fact, most of the predictions are more complicated (most predictions differed by more than one word, and most methods are larger).  Consider that there were 6945 methods out of 90908 in the test set where {\small \texttt{ast-attendgru+FC}} outperformed {\small \texttt{ast-attendgru}} in terms of aggregate BLEU score.  Of these, 5093 (73\%) had words in the reference summary that were in the file context but not the source code of the method; {\small \texttt{ast-attendgru+FC}} correctly used these words in the output predictions for 4369 (86\%).  Put another way, there were 4369/90908 (about 5\%) methods in the test set where {\small \texttt{ast-attendgru}} failed to find the correct word, but {\small \texttt{ast-attendgru+FC}} did find the correct word, and that word was in the file context.  Improvement over this 5\% largely explains the increase in aggregate BLEU score from 18.69 for {\small \texttt{ast-attendgru}} to 19.95 for {\small \texttt{ast-attendgru+FC}} (+6.7\%).  However, in practice, the file context sometimes led the model astray.  Of the 90908 test set methods, {\small \texttt{ast-attendgru}} and {\small \texttt{ast-attendgru+FC}} tied the aggregate BLEU score 78682 times.  For the 12226 times they differed, as mentioned, 6945 times {\small \texttt{ast-attendgru+FC}} outperformed {\small \texttt{ast-attendgru}} while 5281 times it underperformed.  Of these, 2761 (52\%) occurred when a {\small \texttt{ast-attendgru+FC}} picked a word from the file context when the reference did not contain that word.

As a final piece of evidence studying the effect of file context, we perform an ablation study in which we train and test two models in an extreme condition: when \emph{zero} code/text data are available.  Basically we train and test with AST and file context only, code/text sequences are set to zeros.  Ablation studies are common in NMT research to determine whether a given input is benefiting a model~\cite{kuncoro2017recurrent}.  Our study is akin to the challenge experiment proposed by LeClair~\emph{et al.}~\cite{leclair2019neural}, but differs in that our intent is to explore file context rather than simulate code obfuscation.  For brevity, we present only BLEU score values for {\small \texttt{ast-attendgru}} and {\small \texttt{ast-attendgru+FC}}:

\begin{table}[h!]
	\vspace{-0.15cm}
	\begin{tabular}{llllll}
		\multicolumn{1}{r}{BLEU:}             & \multicolumn{1}{c}{A} & \multicolumn{1}{c}{1} & \multicolumn{1}{c}{2} & \multicolumn{1}{c}{3} & \multicolumn{1}{c}{4} \\ \cline{1-1}
		\multicolumn{1}{l|}{ast-attendgru}    & \textbf{8.51}         & 23.37                 & 10.11                 & 5.49                 & 4.05                  \\
		\multicolumn{1}{l|}{ast-attendgru+FC} & \textbf{11.31}        & 27.86                 & 13.04                 & 7.78                  & 5.79                 
	\end{tabular}
	\vspace{-0.19cm}
\end{table}

The {\small \texttt{ast-attendgru}} model had only the AST from which to base predictions.  As shown in the prediction for Example 1 below, it identified that the method sets some value (because it receives a parameter and sets a class property to that parameter's value), but without any text it only predicts an unknown token.  On the other hand, {\small \texttt{ast-attendgru+FC}} guesses terms from the file context such as ``departure airport.''  While not technically a correct prediction, this example illustrates how the file context can serve as an alternative to text information in the source code of the method.  Overall, the aggregate BLEU score increases from 8.51 to 11.31 (33\%).  This difference is almost certainly due to the file context, since the code/text data are ablated and all other details of the model are identical.

\begin{table}[h!]
	\vspace{-0.15cm}
	{\small	
	\begin{tabular}{lm{5cm}} 
		\emph{reference} & sets the intermediate value for this flight  	\\ \hline
		ast-attendgru    & sets the <UNK> value for this <UNK>					\\ 
		ast-attendgru+FC & sets the departure airport value for this flight		\\ 
		
	\end{tabular}
	}
\end{table}

\section{Discussion / Conclusion}
\label{sec:reproducibility}
\vspace{0.1cm}

This paper advances the state of the art by demonstrating how file context can be used to improve neural source code summarization.  The idea that file context includes important clues for understanding subroutines is well-studied in software engineering and especially program comprehension -- it has even been proposed for code summarization~\cite{hill2009automatically, mcburney2016automatic}.  Yet the question is how to make use of those clues in practice.  In a nutshell, our approach is to encode in a recurrent network every subroutine in the same file as the subroutine we are trying to summarize.  
The result is that the model is able to use more words from that context, as in the following examples (for maximum reproducibility, number is method ID in the dataset followed by the reference summary, attention visualizations and detailed explanation are in our online appendix):

\begin{figure}[!h]
	\vspace{-0.6cm}
	{\small
		\begin{tabular}{ll}
			\emph{14624250}    & test of decode nmea method of class org                  \\
			ast-ag.    & test of decode $<$UNK$>$                                 \\
			ast-ag.+FC & test of decode nmea system method of class $<$UNK$>$     \\ \hline
			\emph{51016854}    & returns the weight of the given edge                     \\
			ast-ag.    & returns the weight of the attribute                      \\
			ast-ag.+FC & returns the weight of the given edge                     \\ \hline
			\emph{37563332}    & adds the given rectangle to the collection of polygons   \\
			ast-ag.    & adds the given x y to the collection of values           \\
			ast-ag.+FC & adds the given rectangle to the collection of polygons   \\ \hline
			\emph{37563423}    & returns the height of the font described by the receiver \\
			ast-ag.    & returns the height of the window                         \\
			ast-ag.+FC & returns the height of the font                          
		\end{tabular}
	}
	\vspace{-0.6cm}
\end{figure}

As with all papers, our work carries threats to validity and limitations.  For one, we use a quantitative evaluation methodology, which while in line with almost all related work and which enables us to evaluate the approach over tens of thousands of subroutines, may miss nuances that a qualitative human evaluation would catch.  However, space limitations prevent us from including a thorough discussion of both in a single paper, so we defer a human evaluation for extended work.  Another limitation is that we evaluate only against a Java dataset.  This dataset is the largest available that follows good practice to avoid biases~\cite{leclair2019recommendations}, but caution is advisable when generalizing the work to other languages.  

Still, we demonstrate that an advantage to our approach is that it improves predictions in a way that is orthogonal to existing approaches, so it can be applied to a variety of existing solutions.
A feature of our experiment is that we simplified and reimplemented baselines in order to create a controlled environment for evaluation -- several of these baselines had modifications unrelated to software engineering such as subtoken encoding, ensemble methods, or different RNN architectures, and it was necessary to eliminate these as factors in experimental outcome.  An optimistic sign for future work is that performance reported in this paper could rise further when our approach is combined with these modifications.

\vspace{0.1cm}
\textbf{Reproducibility.}  We release our implementation and supporting scripts via an online appendix / repository:

\vspace{0.0cm}
\textbf{\url{http://www.github.com/Attn-to-FC/Attn-to-FC}}

\vspace{-0.1cm}
\section{Acknowledgments}

This work is supported in part by NSF CCF-1452959 and CCF-1717607. Any opinions, findings, and conclusions expressed herein are the authors and do not necessarily reflect those of the sponsors.


\bibliographystyle{ACM-Reference-Format}
\bibliography{main}


\begin{thebibliography}{47}


\ifx \showCODEN    \undefined \def \showCODEN     #1{\unskip}     \fi
\ifx \showDOI      \undefined \def \showDOI       #1{#1}\fi
\ifx \showISBNx    \undefined \def \showISBNx     #1{\unskip}     \fi
\ifx \showISBNxiii \undefined \def \showISBNxiii  #1{\unskip}     \fi
\ifx \showISSN     \undefined \def \showISSN      #1{\unskip}     \fi
\ifx \showLCCN     \undefined \def \showLCCN      #1{\unskip}     \fi
\ifx \shownote     \undefined \def \shownote      #1{#1}          \fi
\ifx \showarticletitle \undefined \def \showarticletitle #1{#1}   \fi
\ifx \showURL      \undefined \def \showURL       {\relax}        \fi
\providecommand\bibfield[2]{#2}
\providecommand\bibinfo[2]{#2}
\providecommand\natexlab[1]{#1}
\providecommand\showeprint[2][]{arXiv:#2}

\bibitem[\protect\citeauthoryear{Allamanis}{Allamanis}{2019}]%
        {ml4codewebsite}
\bibfield{author}{\bibinfo{person}{Miltos Allamanis}.}
  \bibinfo{year}{2019}\natexlab{}.
\newblock \bibinfo{title}{Machine Learning for Big Code and Naturalness}.
\newblock
\newblock
\urldef\tempurl%
\url{https://ml4code.github.io/papers.html}
\showURL{%
\tempurl}


\bibitem[\protect\citeauthoryear{Allamanis, Barr, Devanbu, and
  Sutton}{Allamanis et~al\mbox{.}}{2017}]%
        {allamanis2017survey}
\bibfield{author}{\bibinfo{person}{Miltiadis Allamanis},
  \bibinfo{person}{Earl~T Barr}, \bibinfo{person}{Premkumar Devanbu}, {and}
  \bibinfo{person}{Charles Sutton}.} \bibinfo{year}{2017}\natexlab{}.
\newblock \showarticletitle{A survey of machine learning for big code and
  naturalness}.
\newblock \bibinfo{journal}{\emph{arXiv preprint arXiv:1709.06182}}
  (\bibinfo{year}{2017}).
\newblock


\bibitem[\protect\citeauthoryear{Allamanis, Brockschmidt, and
  Khademi}{Allamanis et~al\mbox{.}}{2018}]%
        {allamanis2018learning}
\bibfield{author}{\bibinfo{person}{Miltiadis Allamanis}, \bibinfo{person}{Marc
  Brockschmidt}, {and} \bibinfo{person}{Mahmoud Khademi}.}
  \bibinfo{year}{2018}\natexlab{}.
\newblock \showarticletitle{Learning to represent programs with graphs}.
\newblock \bibinfo{journal}{\emph{International Conference on Learning
  Representations}} (\bibinfo{year}{2018}).
\newblock


\bibitem[\protect\citeauthoryear{Alon, Brody, Levy, and Yahav}{Alon
  et~al\mbox{.}}{2019}]%
        {alon2018code2seq}
\bibfield{author}{\bibinfo{person}{Uri Alon}, \bibinfo{person}{Shaked Brody},
  \bibinfo{person}{Omer Levy}, {and} \bibinfo{person}{Eran Yahav}.}
  \bibinfo{year}{2019}\natexlab{}.
\newblock \showarticletitle{code2seq: Generating sequences from structured
  representations of code}.
\newblock \bibinfo{journal}{\emph{International Conference on Learning
  Representations}} (\bibinfo{year}{2019}).
\newblock


\bibitem[\protect\citeauthoryear{Bahdanau, Cho, and Bengio}{Bahdanau
  et~al\mbox{.}}{2014}]%
        {bahdanau2014neural}
\bibfield{author}{\bibinfo{person}{Dzmitry Bahdanau},
  \bibinfo{person}{Kyunghyun Cho}, {and} \bibinfo{person}{Yoshua Bengio}.}
  \bibinfo{year}{2014}\natexlab{}.
\newblock \showarticletitle{Neural machine translation by jointly learning to
  align and translate}.
\newblock \bibinfo{journal}{\emph{arXiv preprint arXiv:1409.0473}}
  (\bibinfo{year}{2014}).
\newblock


\bibitem[\protect\citeauthoryear{Balog, Gaunt, Brockschmidt, Nowozin, and
  Tarlow}{Balog et~al\mbox{.}}{2017}]%
        {balog2017deepcoder}
\bibfield{author}{\bibinfo{person}{M Balog}, \bibinfo{person}{AL Gaunt},
  \bibinfo{person}{M Brockschmidt}, \bibinfo{person}{S Nowozin}, {and}
  \bibinfo{person}{D Tarlow}.} \bibinfo{year}{2017}\natexlab{}.
\newblock \showarticletitle{DeepCoder: Learning to Write Programs}. In
  \bibinfo{booktitle}{\emph{International Conference on Learning
  Representations (ICLR 2017)}}. OpenReview. net.
\newblock


\bibitem[\protect\citeauthoryear{Biggerstaff, Mitbander, and
  Webster}{Biggerstaff et~al\mbox{.}}{1993}]%
        {biggerstaff1993concept}
\bibfield{author}{\bibinfo{person}{Ted~J Biggerstaff},
  \bibinfo{person}{Bharat~G Mitbander}, {and} \bibinfo{person}{Dallas
  Webster}.} \bibinfo{year}{1993}\natexlab{}.
\newblock \showarticletitle{The concept assignment problem in program
  understanding}. In \bibinfo{booktitle}{\emph{Proceedings of the 15th
  international conference on Software Engineering}}. IEEE Computer Society
  Press, \bibinfo{pages}{482--498}.
\newblock


\bibitem[\protect\citeauthoryear{Chen, Firat, Bapna, Johnson, Macherey, Foster,
  Jones, Schuster, Shazeer, Parmar, et~al\mbox{.}}{Chen et~al\mbox{.}}{2018}]%
        {chen2018best}
\bibfield{author}{\bibinfo{person}{Mia~Xu Chen}, \bibinfo{person}{Orhan Firat},
  \bibinfo{person}{Ankur Bapna}, \bibinfo{person}{Melvin Johnson},
  \bibinfo{person}{Wolfgang Macherey}, \bibinfo{person}{George Foster},
  \bibinfo{person}{Llion Jones}, \bibinfo{person}{Mike Schuster},
  \bibinfo{person}{Noam Shazeer}, \bibinfo{person}{Niki Parmar},
  {et~al\mbox{.}}} \bibinfo{year}{2018}\natexlab{}.
\newblock \showarticletitle{The Best of Both Worlds: Combining Recent Advances
  in Neural Machine Translation}. In \bibinfo{booktitle}{\emph{Proceedings of
  the 56th Annual Meeting of the Association for Computational Linguistics
  (Volume 1: Long Papers)}}. \bibinfo{pages}{76--86}.
\newblock


\bibitem[\protect\citeauthoryear{Chen, Wu, and Zaki}{Chen
  et~al\mbox{.}}{2020}]%
        {chen2019reinforcement}
\bibfield{author}{\bibinfo{person}{Yu Chen}, \bibinfo{person}{Lingfei Wu},
  {and} \bibinfo{person}{Mohammed~J Zaki}.} \bibinfo{year}{2020}\natexlab{}.
\newblock \showarticletitle{Reinforcement learning based graph-to-sequence
  model for natural question generation}.
\newblock \bibinfo{journal}{\emph{International Conference on Learning
  Representations}} (\bibinfo{year}{2020}).
\newblock


\bibitem[\protect\citeauthoryear{Chu and Wang}{Chu and Wang}{2018}]%
        {chu2018survey}
\bibfield{author}{\bibinfo{person}{Chenhui Chu} {and} \bibinfo{person}{Rui
  Wang}.} \bibinfo{year}{2018}\natexlab{}.
\newblock \showarticletitle{A Survey of Domain Adaptation for Neural Machine
  Translation}. In \bibinfo{booktitle}{\emph{Proceedings of the 27th
  International Conference on Computational Linguistics}}.
  \bibinfo{pages}{1304--1319}.
\newblock


\bibitem[\protect\citeauthoryear{Cohen and Devanbu}{Cohen and Devanbu}{2018}]%
        {NL4SEAAAI:2018}
\bibfield{author}{\bibinfo{person}{William Cohen} {and} \bibinfo{person}{Prem
  Devanbu}.} \bibinfo{year}{2018}\natexlab{}.
\newblock \bibinfo{title}{Workshop on NLP for Software Engineering}.
\newblock
\newblock
\urldef\tempurl%
\url{https://nl4se.github.io/}
\showURL{%
\tempurl}


\bibitem[\protect\citeauthoryear{Doya}{Doya}{1992}]%
        {doya1992bifurcations}
\bibfield{author}{\bibinfo{person}{Kenji Doya}.}
  \bibinfo{year}{1992}\natexlab{}.
\newblock \showarticletitle{Bifurcations in the learning of recurrent neural
  networks}. In \bibinfo{booktitle}{\emph{[Proceedings] 1992 IEEE International
  Symposium on Circuits and Systems}}, Vol.~\bibinfo{volume}{6}. IEEE,
  \bibinfo{pages}{2777--2780}.
\newblock


\bibitem[\protect\citeauthoryear{Doya}{Doya}{2003}]%
        {doya2003recurrent}
\bibfield{author}{\bibinfo{person}{Kenji Doya}.}
  \bibinfo{year}{2003}\natexlab{}.
\newblock \showarticletitle{Recurrent networks: learning algorithms}.
\newblock \bibinfo{journal}{\emph{The Handbook of Brain Theory and Neural
  Networks,}} (\bibinfo{year}{2003}), \bibinfo{pages}{955--960}.
\newblock


\bibitem[\protect\citeauthoryear{Forward and Lethbridge}{Forward and
  Lethbridge}{2002}]%
        {forward2002relevance}
\bibfield{author}{\bibinfo{person}{Andrew Forward} {and}
  \bibinfo{person}{Timothy~C Lethbridge}.} \bibinfo{year}{2002}\natexlab{}.
\newblock \showarticletitle{The relevance of software documentation, tools and
  technologies: a survey}. In \bibinfo{booktitle}{\emph{Proceedings of the 2002
  ACM symposium on Document engineering}}. ACM, \bibinfo{pages}{26--33}.
\newblock


\bibitem[\protect\citeauthoryear{Gu, Zhang, and Kim}{Gu et~al\mbox{.}}{2018}]%
        {gu2018deep}
\bibfield{author}{\bibinfo{person}{Xiaodong Gu}, \bibinfo{person}{Hongyu
  Zhang}, {and} \bibinfo{person}{Sunghun Kim}.}
  \bibinfo{year}{2018}\natexlab{}.
\newblock \showarticletitle{Deep code search}. In
  \bibinfo{booktitle}{\emph{Proceedings of the 40th International Conference on
  Software Engineering}}. ACM, \bibinfo{pages}{933--944}.
\newblock


\bibitem[\protect\citeauthoryear{Haiduc, Aponte, Moreno, and Marcus}{Haiduc
  et~al\mbox{.}}{2010}]%
        {haiduc2010use}
\bibfield{author}{\bibinfo{person}{Sonia Haiduc}, \bibinfo{person}{Jairo
  Aponte}, \bibinfo{person}{Laura Moreno}, {and} \bibinfo{person}{Andrian
  Marcus}.} \bibinfo{year}{2010}\natexlab{}.
\newblock \showarticletitle{On the use of automated text summarization
  techniques for summarizing source code}. In \bibinfo{booktitle}{\emph{Reverse
  Engineering (WCRE), 2010 17th Working Conference on}}. IEEE,
  \bibinfo{pages}{35--44}.
\newblock


\bibitem[\protect\citeauthoryear{Hellendoorn and Devanbu}{Hellendoorn and
  Devanbu}{2017}]%
        {hellendoorn2017deep}
\bibfield{author}{\bibinfo{person}{Vincent~J Hellendoorn} {and}
  \bibinfo{person}{Premkumar Devanbu}.} \bibinfo{year}{2017}\natexlab{}.
\newblock \showarticletitle{Are deep neural networks the best choice for
  modeling source code?}. In \bibinfo{booktitle}{\emph{Proceedings of the 2017
  11th Joint Meeting on Foundations of Software Engineering}}. ACM,
  \bibinfo{pages}{763--773}.
\newblock


\bibitem[\protect\citeauthoryear{Hill, Pollock, and Vijay-Shanker}{Hill
  et~al\mbox{.}}{2009}]%
        {hill2009automatically}
\bibfield{author}{\bibinfo{person}{Emily Hill}, \bibinfo{person}{Lori Pollock},
  {and} \bibinfo{person}{K Vijay-Shanker}.} \bibinfo{year}{2009}\natexlab{}.
\newblock \showarticletitle{Automatically capturing source code context of
  nl-queries for software maintenance and reuse}. In
  \bibinfo{booktitle}{\emph{Proceedings of the 31st International Conference on
  Software Engineering}}. IEEE Computer Society, \bibinfo{pages}{232--242}.
\newblock


\bibitem[\protect\citeauthoryear{Holmes and Murphy}{Holmes and Murphy}{2005}]%
        {holmes2005using}
\bibfield{author}{\bibinfo{person}{Reid Holmes} {and} \bibinfo{person}{Gail~C
  Murphy}.} \bibinfo{year}{2005}\natexlab{}.
\newblock \showarticletitle{Using structural context to recommend source code
  examples}. In \bibinfo{booktitle}{\emph{Proceedings. 27th International
  Conference on Software Engineering, 2005. ICSE 2005.}} IEEE,
  \bibinfo{pages}{117--125}.
\newblock


\bibitem[\protect\citeauthoryear{Hossain, Sohel, Shiratuddin, and Laga}{Hossain
  et~al\mbox{.}}{2019}]%
        {hossain2019comprehensive}
\bibfield{author}{\bibinfo{person}{MD Hossain}, \bibinfo{person}{Ferdous
  Sohel}, \bibinfo{person}{Mohd~Fairuz Shiratuddin}, {and}
  \bibinfo{person}{Hamid Laga}.} \bibinfo{year}{2019}\natexlab{}.
\newblock \showarticletitle{A comprehensive survey of deep learning for image
  captioning}.
\newblock \bibinfo{journal}{\emph{ACM Computing Surveys (CSUR)}}
  \bibinfo{volume}{51}, \bibinfo{number}{6} (\bibinfo{year}{2019}),
  \bibinfo{pages}{118}.
\newblock


\bibitem[\protect\citeauthoryear{Hu, Li, Xia, Lo, and Jin}{Hu
  et~al\mbox{.}}{2018}]%
        {hu2018deep}
\bibfield{author}{\bibinfo{person}{Xing Hu}, \bibinfo{person}{Ge Li},
  \bibinfo{person}{Xin Xia}, \bibinfo{person}{David Lo}, {and}
  \bibinfo{person}{Zhi Jin}.} \bibinfo{year}{2018}\natexlab{}.
\newblock \showarticletitle{Deep code comment generation}. In
  \bibinfo{booktitle}{\emph{Proceedings of the 26th International Conference on
  Program Comprehension}}. ACM, \bibinfo{pages}{200--210}.
\newblock


\bibitem[\protect\citeauthoryear{Iyer, Konstas, Cheung, and Zettlemoyer}{Iyer
  et~al\mbox{.}}{2016}]%
        {iyer2016summarizing}
\bibfield{author}{\bibinfo{person}{Srinivasan Iyer}, \bibinfo{person}{Ioannis
  Konstas}, \bibinfo{person}{Alvin Cheung}, {and} \bibinfo{person}{Luke
  Zettlemoyer}.} \bibinfo{year}{2016}\natexlab{}.
\newblock \showarticletitle{Summarizing source code using a neural attention
  model}. In \bibinfo{booktitle}{\emph{Proceedings of the 54th Annual Meeting
  of the Association for Computational Linguistics (Volume 1: Long Papers)}},
  Vol.~\bibinfo{volume}{1}. \bibinfo{pages}{2073--2083}.
\newblock


\bibitem[\protect\citeauthoryear{Jiang, Armaly, and McMillan}{Jiang
  et~al\mbox{.}}{2017}]%
        {jiang2017automatically}
\bibfield{author}{\bibinfo{person}{Siyuan Jiang}, \bibinfo{person}{Ameer
  Armaly}, {and} \bibinfo{person}{Collin McMillan}.}
  \bibinfo{year}{2017}\natexlab{}.
\newblock \showarticletitle{Automatically generating commit messages from diffs
  using neural machine translation}. In \bibinfo{booktitle}{\emph{Proceedings
  of the 32nd IEEE/ACM International Conference on Automated Software
  Engineering}}. IEEE Press, \bibinfo{pages}{135--146}.
\newblock


\bibitem[\protect\citeauthoryear{Kramer}{Kramer}{1999}]%
        {kramer1999api}
\bibfield{author}{\bibinfo{person}{Douglas Kramer}.}
  \bibinfo{year}{1999}\natexlab{}.
\newblock \showarticletitle{API documentation from source code comments: a case
  study of Javadoc}. In \bibinfo{booktitle}{\emph{Proceedings of the 17th
  annual international conference on Computer documentation}}. ACM,
  \bibinfo{pages}{147--153}.
\newblock


\bibitem[\protect\citeauthoryear{Kuncoro, Ballesteros, Kong, Dyer, Neubig, and
  Smith}{Kuncoro et~al\mbox{.}}{2017}]%
        {kuncoro2017recurrent}
\bibfield{author}{\bibinfo{person}{Adhiguna Kuncoro}, \bibinfo{person}{Miguel
  Ballesteros}, \bibinfo{person}{Lingpeng Kong}, \bibinfo{person}{Chris Dyer},
  \bibinfo{person}{Graham Neubig}, {and} \bibinfo{person}{Noah~A Smith}.}
  \bibinfo{year}{2017}\natexlab{}.
\newblock \showarticletitle{What Do Recurrent Neural Network Grammars Learn
  About Syntax?}. In \bibinfo{booktitle}{\emph{Proceedings of the 15th
  Conference of the European Chapter of the Association for Computational
  Linguistics: Volume 1, Long Papers}}. \bibinfo{pages}{1249--1258}.
\newblock


\bibitem[\protect\citeauthoryear{LeClair, Jiang, and McMillan}{LeClair
  et~al\mbox{.}}{2019}]%
        {leclair2019neural}
\bibfield{author}{\bibinfo{person}{Alexander LeClair}, \bibinfo{person}{Siyuan
  Jiang}, {and} \bibinfo{person}{Collin McMillan}.}
  \bibinfo{year}{2019}\natexlab{}.
\newblock \showarticletitle{A neural model for generating natural language
  summaries of program subroutines}. In \bibinfo{booktitle}{\emph{Proceedings
  of the 41st International Conference on Software Engineering}}. IEEE Press,
  \bibinfo{pages}{795--806}.
\newblock


\bibitem[\protect\citeauthoryear{LeClair and McMillan}{LeClair and
  McMillan}{2019}]%
        {leclair2019recommendations}
\bibfield{author}{\bibinfo{person}{Alexander LeClair} {and}
  \bibinfo{person}{Collin McMillan}.} \bibinfo{year}{2019}\natexlab{}.
\newblock \showarticletitle{Recommendations for Datasets for Source Code
  Summarization}. In \bibinfo{booktitle}{\emph{Proceedings of the 2019
  Conference of the North American Chapter of the Association for Computational
  Linguistics: Human Language Technologies, Volume 1 (Long and Short Papers)}}.
  \bibinfo{pages}{3931--3937}.
\newblock


\bibitem[\protect\citeauthoryear{Lin}{Lin}{2004}]%
        {lin2004rouge}
\bibfield{author}{\bibinfo{person}{Chin-Yew Lin}.}
  \bibinfo{year}{2004}\natexlab{}.
\newblock \showarticletitle{Rouge: A package for automatic evaluation of
  summaries}.
\newblock \bibinfo{journal}{\emph{Text Summarization Branches Out}}
  (\bibinfo{year}{2004}).
\newblock


\bibitem[\protect\citeauthoryear{Lopes, Bajracharya, Ossher, and Baldi}{Lopes
  et~al\mbox{.}}{2010}]%
        {Lopes+Bajracharya+Ossher+Baldi:2010}
\bibfield{author}{\bibinfo{person}{C. Lopes}, \bibinfo{person}{S. Bajracharya},
  \bibinfo{person}{J. Ossher}, {and} \bibinfo{person}{P. Baldi}.}
  \bibinfo{year}{2010}\natexlab{}.
\newblock \bibinfo{title}{{UCI} Source Code Data Sets}.
\newblock
\newblock
\urldef\tempurl%
\url{http://www.ics.uci.edu/$\sim$lopes/datasets/}
\showURL{%
\tempurl}


\bibitem[\protect\citeauthoryear{McBurney and McMillan}{McBurney and
  McMillan}{2016}]%
        {mcburney2016automatic}
\bibfield{author}{\bibinfo{person}{Paul~W McBurney} {and}
  \bibinfo{person}{Collin McMillan}.} \bibinfo{year}{2016}\natexlab{}.
\newblock \showarticletitle{Automatic source code summarization of context for
  java methods}.
\newblock \bibinfo{journal}{\emph{IEEE Transactions on Software Engineering}}
  \bibinfo{volume}{42}, \bibinfo{number}{2} (\bibinfo{year}{2016}),
  \bibinfo{pages}{103--119}.
\newblock


\bibitem[\protect\citeauthoryear{Nallapati, Zhou, dos Santos, Gulcehre, and
  Xiang}{Nallapati et~al\mbox{.}}{2016}]%
        {nallapati2016abstractive}
\bibfield{author}{\bibinfo{person}{Ramesh Nallapati}, \bibinfo{person}{Bowen
  Zhou}, \bibinfo{person}{Cicero dos Santos}, \bibinfo{person}{Caglar
  Gulcehre}, {and} \bibinfo{person}{Bing Xiang}.}
  \bibinfo{year}{2016}\natexlab{}.
\newblock \showarticletitle{Abstractive Text Summarization using
  Sequence-to-sequence RNNs and Beyond}. In
  \bibinfo{booktitle}{\emph{Proceedings of The 20th SIGNLL Conference on
  Computational Natural Language Learning}}. \bibinfo{pages}{280--290}.
\newblock


\bibitem[\protect\citeauthoryear{Nazar, Hu, and Jiang}{Nazar
  et~al\mbox{.}}{2016}]%
        {nazar2016summarizing}
\bibfield{author}{\bibinfo{person}{Najam Nazar}, \bibinfo{person}{Yan Hu},
  {and} \bibinfo{person}{He Jiang}.} \bibinfo{year}{2016}\natexlab{}.
\newblock \showarticletitle{Summarizing software artifacts: A literature
  review}.
\newblock \bibinfo{journal}{\emph{Journal of Computer Science and Technology}}
  \bibinfo{volume}{31}, \bibinfo{number}{5} (\bibinfo{year}{2016}),
  \bibinfo{pages}{883--909}.
\newblock


\bibitem[\protect\citeauthoryear{Oda, Fudaba, Neubig, Hata, Sakti, Toda, and
  Nakamura}{Oda et~al\mbox{.}}{2015}]%
        {oda2015learning}
\bibfield{author}{\bibinfo{person}{Yusuke Oda}, \bibinfo{person}{Hiroyuki
  Fudaba}, \bibinfo{person}{Graham Neubig}, \bibinfo{person}{Hideaki Hata},
  \bibinfo{person}{Sakriani Sakti}, \bibinfo{person}{Tomoki Toda}, {and}
  \bibinfo{person}{Satoshi Nakamura}.} \bibinfo{year}{2015}\natexlab{}.
\newblock \showarticletitle{Learning to generate pseudo-code from source code
  using statistical machine translation (t)}. In
  \bibinfo{booktitle}{\emph{Automated Software Engineering (ASE), 2015 30th
  IEEE/ACM International Conference on}}. IEEE, \bibinfo{pages}{574--584}.
\newblock


\bibitem[\protect\citeauthoryear{Olden and Jackson}{Olden and Jackson}{2002}]%
        {olden2002illuminating}
\bibfield{author}{\bibinfo{person}{Julian~D Olden} {and}
  \bibinfo{person}{Donald~A Jackson}.} \bibinfo{year}{2002}\natexlab{}.
\newblock \showarticletitle{Illuminating the “black box”: a randomization
  approach for understanding variable contributions in artificial neural
  networks}.
\newblock \bibinfo{journal}{\emph{Ecological modelling}} \bibinfo{volume}{154},
  \bibinfo{number}{1-2} (\bibinfo{year}{2002}), \bibinfo{pages}{135--150}.
\newblock


\bibitem[\protect\citeauthoryear{Papineni, Roukos, Ward, and Zhu}{Papineni
  et~al\mbox{.}}{2002}]%
        {Papineni:2002:BMA:1073083.1073135}
\bibfield{author}{\bibinfo{person}{Kishore Papineni}, \bibinfo{person}{Salim
  Roukos}, \bibinfo{person}{Todd Ward}, {and} \bibinfo{person}{Wei-Jing Zhu}.}
  \bibinfo{year}{2002}\natexlab{}.
\newblock \showarticletitle{BLEU: A Method for Automatic Evaluation of Machine
  Translation}. In \bibinfo{booktitle}{\emph{Proceedings of the 40th Annual
  Meeting on Association for Computational Linguistics}}
  \emph{(\bibinfo{series}{ACL '02})}. \bibinfo{publisher}{Association for
  Computational Linguistics}, \bibinfo{address}{Stroudsburg, PA, USA},
  \bibinfo{pages}{311--318}.
\newblock
\urldef\tempurl%
\url{https://doi.org/10.3115/1073083.1073135}
\showDOI{\tempurl}


\bibitem[\protect\citeauthoryear{Pouyanfar, Sadiq, Yan, Tian, Tao, Reyes, Shyu,
  Chen, and Iyengar}{Pouyanfar et~al\mbox{.}}{2018}]%
        {pouyanfar2018survey}
\bibfield{author}{\bibinfo{person}{Samira Pouyanfar}, \bibinfo{person}{Saad
  Sadiq}, \bibinfo{person}{Yilin Yan}, \bibinfo{person}{Haiman Tian},
  \bibinfo{person}{Yudong Tao}, \bibinfo{person}{Maria~Presa Reyes},
  \bibinfo{person}{Mei-Ling Shyu}, \bibinfo{person}{Shu-Ching Chen}, {and}
  \bibinfo{person}{SS Iyengar}.} \bibinfo{year}{2018}\natexlab{}.
\newblock \showarticletitle{A survey on deep learning: Algorithms, techniques,
  and applications}.
\newblock \bibinfo{journal}{\emph{ACM Computing Surveys (CSUR)}}
  \bibinfo{volume}{51}, \bibinfo{number}{5} (\bibinfo{year}{2018}),
  \bibinfo{pages}{92}.
\newblock


\bibitem[\protect\citeauthoryear{Shrestha and Mahmood}{Shrestha and
  Mahmood}{2019}]%
        {shrestha2019review}
\bibfield{author}{\bibinfo{person}{Ajay Shrestha} {and} \bibinfo{person}{Ausif
  Mahmood}.} \bibinfo{year}{2019}\natexlab{}.
\newblock \showarticletitle{Review of Deep Learning Algorithms and
  Architectures}.
\newblock \bibinfo{journal}{\emph{IEEE Access}}  \bibinfo{volume}{7}
  (\bibinfo{year}{2019}), \bibinfo{pages}{53040--53065}.
\newblock


\bibitem[\protect\citeauthoryear{Song, Sun, Wang, and Yan}{Song
  et~al\mbox{.}}{2019}]%
        {song2019survey}
\bibfield{author}{\bibinfo{person}{Xiaotao Song}, \bibinfo{person}{Hailong
  Sun}, \bibinfo{person}{Xu Wang}, {and} \bibinfo{person}{Jiafei Yan}.}
  \bibinfo{year}{2019}\natexlab{}.
\newblock \showarticletitle{A Survey of Automatic Generation of Source Code
  Comments: Algorithms and Techniques}.
\newblock \bibinfo{journal}{\emph{IEEE Access}} (\bibinfo{year}{2019}).
\newblock


\bibitem[\protect\citeauthoryear{Sridhara, Hill, Muppaneni, Pollock, and
  Vijay-Shanker}{Sridhara et~al\mbox{.}}{2010}]%
        {sridhara2010towards}
\bibfield{author}{\bibinfo{person}{Giriprasad Sridhara}, \bibinfo{person}{Emily
  Hill}, \bibinfo{person}{Divya Muppaneni}, \bibinfo{person}{Lori Pollock},
  {and} \bibinfo{person}{K Vijay-Shanker}.} \bibinfo{year}{2010}\natexlab{}.
\newblock \showarticletitle{Towards automatically generating summary comments
  for java methods}. In \bibinfo{booktitle}{\emph{Proceedings of the IEEE/ACM
  international conference on Automated software engineering}}. ACM,
  \bibinfo{pages}{43--52}.
\newblock


\bibitem[\protect\citeauthoryear{Sridhara, Pollock, and Vijay-Shanker}{Sridhara
  et~al\mbox{.}}{2011}]%
        {sridhara2011automatically}
\bibfield{author}{\bibinfo{person}{Giriprasad Sridhara}, \bibinfo{person}{Lori
  Pollock}, {and} \bibinfo{person}{K Vijay-Shanker}.}
  \bibinfo{year}{2011}\natexlab{}.
\newblock \showarticletitle{Automatically detecting and describing high level
  actions within methods}. In \bibinfo{booktitle}{\emph{Proceedings of the 33rd
  International Conference on Software Engineering}}. ACM,
  \bibinfo{pages}{101--110}.
\newblock


\bibitem[\protect\citeauthoryear{Vaswani, Shazeer, Parmar, Uszkoreit, Jones,
  Gomez, Kaiser, and Polosukhin}{Vaswani et~al\mbox{.}}{2017}]%
        {vaswani2017attention}
\bibfield{author}{\bibinfo{person}{Ashish Vaswani}, \bibinfo{person}{Noam
  Shazeer}, \bibinfo{person}{Niki Parmar}, \bibinfo{person}{Jakob Uszkoreit},
  \bibinfo{person}{Llion Jones}, \bibinfo{person}{Aidan~N Gomez},
  \bibinfo{person}{{\L}ukasz Kaiser}, {and} \bibinfo{person}{Illia
  Polosukhin}.} \bibinfo{year}{2017}\natexlab{}.
\newblock \showarticletitle{Attention is all you need}. In
  \bibinfo{booktitle}{\emph{Advances in neural information processing
  systems}}. \bibinfo{pages}{5998--6008}.
\newblock


\bibitem[\protect\citeauthoryear{Wan, Zhao, Yang, Xu, Ying, Wu, and Yu}{Wan
  et~al\mbox{.}}{2018}]%
        {wan2018improving}
\bibfield{author}{\bibinfo{person}{Yao Wan}, \bibinfo{person}{Zhou Zhao},
  \bibinfo{person}{Min Yang}, \bibinfo{person}{Guandong Xu},
  \bibinfo{person}{Haochao Ying}, \bibinfo{person}{Jian Wu}, {and}
  \bibinfo{person}{Philip~S Yu}.} \bibinfo{year}{2018}\natexlab{}.
\newblock \showarticletitle{Improving automatic source code summarization via
  deep reinforcement learning}. In \bibinfo{booktitle}{\emph{Proceedings of the
  33rd ACM/IEEE International Conference on Automated Software Engineering}}.
  ACM, \bibinfo{pages}{397--407}.
\newblock


\bibitem[\protect\citeauthoryear{Xu, Wu, Wang, Feng, Witbrock, and Sheinin}{Xu
  et~al\mbox{.}}{2018a}]%
        {xu2018graph2seq}
\bibfield{author}{\bibinfo{person}{Kun Xu}, \bibinfo{person}{Lingfei Wu},
  \bibinfo{person}{Zhiguo Wang}, \bibinfo{person}{Yansong Feng},
  \bibinfo{person}{Michael Witbrock}, {and} \bibinfo{person}{Vadim Sheinin}.}
  \bibinfo{year}{2018}\natexlab{a}.
\newblock \showarticletitle{Graph2seq: Graph to sequence learning with
  attention-based neural networks}.
\newblock \bibinfo{journal}{\emph{arXiv preprint arXiv:1804.00823}}
  (\bibinfo{year}{2018}).
\newblock


\bibitem[\protect\citeauthoryear{Xu, Wu, Wang, Yu, Chen, and Sheinin}{Xu
  et~al\mbox{.}}{2018b}]%
        {xu2018exploiting}
\bibfield{author}{\bibinfo{person}{Kun Xu}, \bibinfo{person}{Lingfei Wu},
  \bibinfo{person}{Zhiguo Wang}, \bibinfo{person}{Mo Yu},
  \bibinfo{person}{Liwei Chen}, {and} \bibinfo{person}{Vadim Sheinin}.}
  \bibinfo{year}{2018}\natexlab{b}.
\newblock \showarticletitle{Exploiting rich syntactic information for semantic
  parsing with graph-to-sequence model}.
\newblock \bibinfo{journal}{\emph{Conference on Empirical Methods in Natural
  Language Processing}} (\bibinfo{year}{2018}).
\newblock


\bibitem[\protect\citeauthoryear{Young, Hazarika, Poria, and Cambria}{Young
  et~al\mbox{.}}{2018}]%
        {young2018recent}
\bibfield{author}{\bibinfo{person}{Tom Young}, \bibinfo{person}{Devamanyu
  Hazarika}, \bibinfo{person}{Soujanya Poria}, {and} \bibinfo{person}{Erik
  Cambria}.} \bibinfo{year}{2018}\natexlab{}.
\newblock \showarticletitle{Recent trends in deep learning based natural
  language processing}.
\newblock \bibinfo{journal}{\emph{ieee Computational intelligenCe magazine}}
  \bibinfo{volume}{13}, \bibinfo{number}{3} (\bibinfo{year}{2018}),
  \bibinfo{pages}{55--75}.
\newblock


\bibitem[\protect\citeauthoryear{Yu, Lam, Chen, Li, Xie, and Wang}{Yu
  et~al\mbox{.}}{2019}]%
        {yu2019neural}
\bibfield{author}{\bibinfo{person}{Hao Yu}, \bibinfo{person}{Wing Lam},
  \bibinfo{person}{Long Chen}, \bibinfo{person}{Ge Li}, \bibinfo{person}{Tao
  Xie}, {and} \bibinfo{person}{Qianxiang Wang}.}
  \bibinfo{year}{2019}\natexlab{}.
\newblock \showarticletitle{Neural detection of semantic code clones via
  tree-based convolution}. In \bibinfo{booktitle}{\emph{Proceedings of the 27th
  International Conference on Program Comprehension}}. IEEE Press,
  \bibinfo{pages}{70--80}.
\newblock


\bibitem[\protect\citeauthoryear{Zhang, Wang, Zhang, Sun, Wang, and Liu}{Zhang
  et~al\mbox{.}}{2019}]%
        {zhang2019novel}
\bibfield{author}{\bibinfo{person}{Jian Zhang}, \bibinfo{person}{Xu Wang},
  \bibinfo{person}{Hongyu Zhang}, \bibinfo{person}{Hailong Sun},
  \bibinfo{person}{Kaixuan Wang}, {and} \bibinfo{person}{Xudong Liu}.}
  \bibinfo{year}{2019}\natexlab{}.
\newblock \showarticletitle{A novel neural source code representation based on
  abstract syntax tree}. In \bibinfo{booktitle}{\emph{Proceedings of the 41st
  International Conference on Software Engineering}}. IEEE Press,
  \bibinfo{pages}{783--794}.
\newblock


\end{thebibliography}

\end{document}